\documentclass[paper]{JHEP3}
\usepackage{cite}
\usepackage{amssymb,amsmath}
\usepackage{graphicx}
\usepackage{booktabs}
\graphicspath{{figures/}}

\def\PR1{
   \parbox[h]{0.15\textwidth}{\includegraphics[width=0.15\textwidth]
{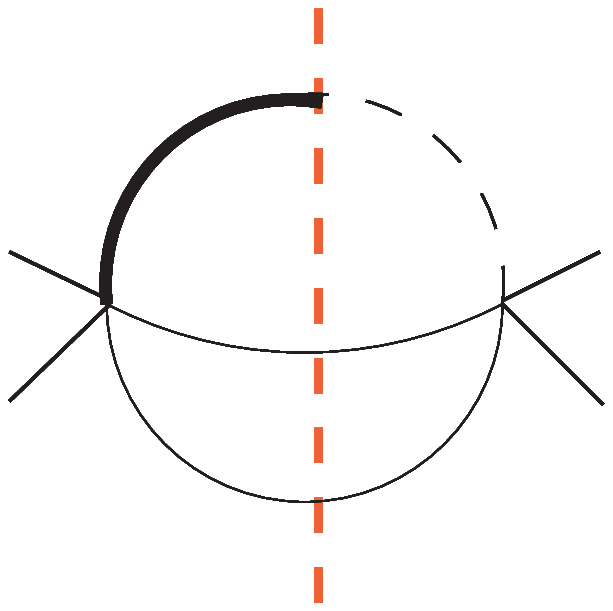}}
}

\newcommand{\I}[1]{
   \parbox[h]{0.15\textwidth}{\includegraphics[width=0.15\textwidth]
{I#1.eps}}
}

\def\ep{\varepsilon}
\def\d{{\rm d}}

\def\dd{\mathcal{D}}

\def\l({\left(}
\def\r){\right)}
\def\G{\hbox{G}}
\def\H{\hbox{H}}
\def\g{\hbox{g}}
\def\dd{\partial}

\newcommand{\abs}[1]{\left| #1 \right|}

\newcommand{\cross}[1]{ \widehat{ #1 } }
\newcommand{ \M }{ \mathcal{M} }
\newcommand{ \q }{ q }
\newcommand{ \qb }{ \overline{q} }
\newcommand{ \gluon }{ g }
\newcommand{ \qp }{ q' }
\newcommand{ \qpb }{ \overline{q}' }
\newcommand{ \qbp }{ \qpb }
\newcommand{ \limfreemark }{*}

\newcommand{ \imom }[1]{ p_{#1} }
\newcommand{ \pa }{ \imom{1} }
\newcommand{ \pb }{ \imom{2} }
\newcommand{ \fmom }[1]{ k_{#1} }
\newcommand{ \fmomenta }{ \fmom{1}, \ldots, \fmom{m+2} }
\newcommand{ \psmomenta }{ \fmomenta ; \pa, \pb }
\newcommand{ \psmomentaini }{ \pa, \pb; \fmomenta }
\newcommand{ \dmom }[1]{ \left[ \d \fmom{#1} \right] }

\newcommand{ \reduced }[1]{ \tilde{ #1 } }
\newcommand{ \fmomred }[1]{ \reduced{ \fmom{} }_{#1} }
\newcommand{ \fmomentared }{ \fmomred{1}, \dotsc, \fmomred{i}, \fmomred{l}, \dotsc, \fmomred{m+2} }
\newcommand{ \xa }{ x_1 }
\newcommand{ \xb }{ x_2 }
\newcommand{ \xahat }{ \hat{x}_1 }
\newcommand{ \xbhat }{ \hat{x}_2 }

\newcommand{ \order }[1]{ \mathcal{O} \left( #1 \right) }

\title{Antenna subtraction at NNLO with hadronic initial states:
double real radiation for initial-initial configurations 
with two quark flavours}

\author{Radja Boughezal \\ High Energy Physics Division, Argonne National
Laboratory, Argonne, IL 60439, USA}

\author{Aude Gehrmann-De Ridder, Mathias Ritzmann\\Institute for Theoretical 
Physics, ETH, CH-8093 Z\"urich, Switzerland}


\abstract{The antenna subtraction formalism allows to 
calculate QCD corrections to jet observables. Within this formalism, 
the subtraction terms are constructed using antenna functions describing 
all unresolved radiation between a pair of hard radiator partons. 
In this paper, we focus on the subtraction terms for double real radiation 
contributions to jet observables in hadron-hadron collisions evaluated at NNLO. 
An essential ingredient to these subtraction terms 
are the four-parton antenna functions with both radiators in the initial 
state. We outline the construction of the double real subtraction terms, 
classify all relevant antenna functions and describe their integration 
over the relevant antenna phase space. 
For the initial-initial antenna functions with two quark flavours, 
we derive the phase space master integrals and obtain the integrated antennae.
}

\preprint{ ANL-HEP-PR-10-58}

\allowdisplaybreaks[4]

\begin{document}

\section{Introduction}
Jet production  processes constitute an important tool 
for precision studies due to their large
cross sections at high energy colliders.
Reliable theoretical predictions for these observables require 
the calculation of at least the next-to-leading order 
QCD corrections.
For these observables, the inclusive cross section 
with two incoming hadrons $H_1,H_2$ can be written as  
\begin{equation}
\d \sigma = \sum_{a,b} \int 
\d \xi_1\d \xi_2\, f_{a/1}(\xi_1) \,f_{b/2}(\xi_2)
\, \d \hat{\sigma}_{ab}(\xi_1H_1,\xi_2H_2)\;, 
\label{eq:part}
\end{equation}
where $\xi_1$ and $\xi_2$ are the momentum fractions of the 
partons of species $a$ and $b$ in the incoming hadrons, $f$ being the 
corresponding parton distribution functions and  
$\d \hat{\sigma}_{ab}(\xi_1H_1,\xi_2H_2)$ is the parton-level 
scattering cross section for incoming partons $a$ and $b$. 

The partonic cross section ${\rm d }\hat{\sigma}_{ab}$  
has a perturbative expansion in the strong coupling $\alpha_{s}$
such that theoretical predictions for a hadronic process 
at a given order in $\alpha_{s}$ are obtained 
when all partonic channels 
contributing to that order of the partonic cross section 
are summed and convoluted with the appropriate parton 
distribution 
functions as in eq.~(\ref{eq:part}).

In general, beyond the leading order, each partonic channel 
contains both ultraviolet and infrared (soft and collinear) divergences.
The ultraviolet poles are removed by renormalisation in each channel.
Collinear poles originating from the radiation of initial state partons 
are cancelled by mass factorisation counterterms and absorbed 
in the parton distribution functions.
The remaining soft and collinear poles cancel among each other when all  
partonic contributions are summed over \cite{KLN}.      
As jet observables 
depend in a nontrivial manner on the experimental criteria used
to define them, they can only be calculated numerically.
The computation of hadronic observables including higher order corrections   
therefore requires a systematic procedure to cancel 
infrared singularities among different partonic channels 
before any numerical computation of the observable can be performed.

For the task of next-to-leading 
order (NLO) calculations, the infrared divergences present in real 
radiation contributions can be systematically extracted by 
process-in\-de\-pen\-dent procedures, called subtraction methods. 
The purpose of any subtraction method at NLO is to provide 
a subtraction term which has the same 
singular behaviour as the real radiation squared matrix element and
is sufficiently simple to be integrated analytically over the radiation phase space 
which has been factorised from the $(m+1)$-particle phase space.
The actual form of this subtraction term 
depends on the subtraction formalism used. Several successful subtraction formalisms have been proposed 
in the literature \cite{Catani:1996vz,Frixione:1995ms,Nagy:1996bz,Frixione:1997np,Somogyi:2006cz}.
Most notably, the FKS \cite{Frixione:1995ms} subtraction by Frixione, Kunszt and Signer 
and the dipole formalism of Catani and Seymour \cite{Catani:1996vz} 
have been implemented in an automated way, the former in \cite{Frederix:2009yq}, the latter in \cite{Gleisberg:2007md,Seymour:2008mu,Hasegawa:2008ae,Hasegawa:2009tx,Frederix:2008hu,Czakon:2009ss}.
The major challenge for NLO calculations is 
the computation of one-loop amplitudes for multiparticle processes. 
The evaluations of $2 \to 4$ processes at the next-to-leading order  
represent the current frontier \cite{Berger:2008sj,Giele:2008bc,Ossola:2007ax,Binoth:2008uq}.

Nevertheless, for some hadronic processes, in particular 
$2 \to1$ or $2 \to 2$ scattering processes 
such as Drell-Yan, Higgs production, dijet production, 
vector-boson plus jet, 
vector-boson pair production or heavy quark pair production, 
the  accuracy of the next-to-leading order 
predictions is not sufficient to match the anticipated  
experimental accuracy, expected to reach the order 
of a few percent or better.
Accurate precision studies enabling the extraction of fundamental 
parameters of the theory will require that the theoretical predictions 
have the same precision. Those need therefore to be evaluated up 
to the next-to-next-to-leading order (NNLO) 
in perturbative QCD\,. \\

The calculation of observables with $m$ jets in addition to other objects (like for example vector bosons) at the NNLO requires
three distinct contributions: the double real radiation 
${\rm{d}}\hat\sigma_{NNLO}^R$ with $(m+2)$ final state partons, 
the mixed real-virtual radiation ${\rm{d}}\hat\sigma_{NNLO}^{V,1}$ 
with $(m+1)$ final state partons and the 2-loop  
virtual radiation ${\rm{d}}\hat\sigma_{NNLO}^{V,2}$, with $m$ final state partons.
Those build the NNLO cross section which is given by
\begin{eqnarray}
{\rm d}\hat\sigma_{NNLO}&=&\int_{{\rm{d}}\Phi_{m+2}} {\rm{d}}\hat\sigma_{NNLO}^R 
+\int_{{\rm{d}}\Phi_{m+1}} {\rm{d}}\hat\sigma_{NNLO}^{V,1} 
+\int_{{\rm{d}}\Phi_m}{\rm{d}}\hat\sigma_{NNLO}^{V,2}.
\end{eqnarray}

The individual contributions in the $m$-, $(m+1)$- and $(m+2)$-parton final states are all separately infrared divergent. After renormalisation 
and factorisation, their sum is finite, though. 
For most massless jet observables of
phenomenological interest, the two-loop matrix elements have been
computed some time ago, while the one-loop matrix elements 
are usually known from calculations of NLO
corrections to $(m+1)$-jet production~\cite{mcfm}\,.
The one-loop and two-loop matrix-elements contain explicit 
infrared divergences from the loop integration. Those cancel with 
divergences which are implicit in the real radiation for $(m+1)$- and $(m+2)$-parton 
processes. These real radiation divergences become explicit 
only once the phase space integration is carried out. 
The main issue of these calculations 
is therefore to find a method to extract and cancel the 
infrared divergences among these three contributions in order 
to finally evaluate numerically the finite remainders  
to obtain the NNLO contribution to the cross section. 

Like at NLO, a subtraction formalism is needed in order to extract 
the infrared divergences from these contributions.
At parton level, the general form of the cross section for an $m$-particle final 
state at NNLO including subtraction terms is given by~\cite{GehrmannDeRidder:2005cm}:

\begin{eqnarray}
{\rm d}\hat\sigma_{NNLO}&=&\int_{{\rm{d}}\Phi_{m+2}}\left({\rm{d}}\hat\sigma_{NNLO}^R-{\rm{d}}\hat\sigma_{NNLO}^S\right)
+\int_{{\rm{d}}\Phi_{m+2}}{\rm{d}}\hat\sigma_{NNLO}^S\nonumber\\
&+&\int_{{\rm{d}}\Phi_{m+1}}\left({\rm{d}}\hat\sigma_{NNLO}^{V,1}-{\rm{d}}\hat\sigma_{NNLO}^{V S,1}\right)
+\int_{{\rm{d}}\Phi_{m+1}}{\rm{d}}\hat\sigma_{NNLO}^{V S,1}
+\int_{{\rm{d}}\Phi_{m+1}}{\rm{d}}\hat\sigma_{NNLO}^{MF,1}\nonumber\\
&+&\int_{{\rm{d}}\Phi_m}{\rm{d}}\hat\sigma_{NNLO}^{V,2}
+\int_{{\rm{d}}\Phi_m}{\rm{d}}\hat\sigma_{NNLO}^{MF,2}.
\label{eq:sigNNLO}
\end{eqnarray}

Here, ${\rm d} \hat\sigma^{S}_{NNLO}$ denotes the subtraction term for the $(m+2)$-parton final state which behaves like the double real radiation contribution
${\rm d} \hat\sigma^{R}_{NNLO}$ in all singular limits. 
Likewise, ${\rm d} \hat\sigma^{VS,1}_{NNLO}$ is the one-loop virtual subtraction 
term 
coinciding with the one-loop $(m+1)$-particle contribution ${\rm d} \hat\sigma^{V,1}_{NNLO}$ in all singular limits. 
The two-loop correction 
to the $m$-parton final state is denoted by ${\rm d}\hat\sigma^{V,2}_{NNLO}$.  In addition, when there are partons in the initial state, 
there are two mass factorisation contributions, 
${\rm d}\hat\sigma^{MF,1}_{NNLO}$ and ${\rm d}\hat\sigma^{MF,2}_{NNLO}$, for the $(m+1)$- and $m$-particle final states respectively.
Like at the next-to-leading order level, 
the subtraction terms are needed in their unintegrated as well
as in their integrated forms.

There have been several approaches to build a general 
subtraction scheme at NNLO \cite{GehrmannDeRidder:2005cm, Weinzierl:2003fx,Frixione:2004is,Somogyi:2005xz,Somogyi:2006da,Somogyi:2006db,Somogyi:2008fc,Aglietti:2008fe,Somogyi:2009ri,Bolzoni:2009ye,Czakon:2010td}. 
Another subtraction scheme, the $q_{T}$-subtraction formalism 
has been proposed in~\cite{Catani:2007vq}. 
It has been applied to evaluate observables 
related to processes with colourless high mass final states \cite{Catani:2007vq,Grazzini:2008tf,Catani:2009sm,Catani:2010en}.
In addition, there is a completely independent approach 
called sector decomposition which relies on a systematic expansion of the integrals in distributions followed by a purely numerical 
integration. It has been developed for virtual~\cite{Binoth:2000ps,Binoth:2003ak,Heinrich:2008si} 
and real radiation~\cite{Heinrich:2002rc,Anastasiou:2003gr, Binoth:2004jv, Heinrich:2006ku} corrections at
NNLO, and applied to several observables 
already~\cite{Anastasiou:2004qd,Anastasiou:2004xq,Anastasiou:2005qj,Melnikov:2006di}.

We will follow the NNLO antenna subtraction method which was derived 
in \cite{GehrmannDeRidder:2005cm} for decays of a colourless initial state into 
massless final state partons. This formalism has been 
applied in the computation of NNLO corrections to three-jet production in electron-positron annihilation 
\cite{GehrmannDeRidder:2007jk,GehrmannDeRidder:2008ug,Weinzierl:2008iv,Weinzierl:2009nz} and related event shapes 
\cite{GehrmannDeRidder:2007bj,GehrmannDeRidder:2007hr,GehrmannDeRidder:2009dp,Weinzierl:2009ms,Weinzierl:2009yz}, 
which were subsequently used in precision determinations 
of the strong coupling constant \cite{Dissertori:2007xa,Dissertori:2009ik,Dissertori:2009qa,Bethke:2008hf,Gehrmann:2009eh}.  

For processes with initial-state partons,
the antenna subtraction formalism has been so far fully worked out 
only to NLO in \cite{Daleo:2006xa}.
It has been extended to NNLO for processes 
involving one initial state parton relevant for electron-proton 
scattering in ~\cite{Daleo:2009yj}  
while an extension of the formalism to include two initial state 
hadrons at NNLO is under construction \cite{Boughezal:2010ty,Joao}. 
An essential step towards this aim is performed in \cite{Joao} where 
an explicit derivation of the subtraction terms needed 
for the double real contributions to the six-gluon process 
is presented.
The general structure of the unintegrated subtraction terms  
relevant for the double real contributions to any hadronic observables 
evaluated at NNLO is presented there as well. 

In this paper, we will focus on the integrated form 
of the subtraction term relevant for double real radiation 
for processes involving two partons in the initial state. 
In the antenna subtraction formalism, 
the subtraction terms are built with so-called antenna functions. 
The latter describe all unresolved partonic radiation 
off a hard pair of colour-ordered partons, the radiators.
Depending on where the two hard radiators are located, three cases need 
to be distinguished: both radiators are in the final state 
(final-final), only one radiator parton is in the initial state 
(initial-final) or both radiator partons are in the initial state 
(initial-initial). The subtraction terms and the antenna functions 
building them are separated corresponding to these three cases. 
In the most general hadronic process 
(two partons in the initial state, two or more partons in the final state), 
all three configurations have to be taken into account.

As discussed in \cite{GehrmannDeRidder:2005cm,Daleo:2009yj,Joao} 
the subtraction terms 
are separated according to the colour connection of the unresolved partons.  
In this paper, we will specialise on the subtraction term 
for the case where two unresolved 
(soft or collinear) partons are colour connected to the two incoming partons.  
This is indeed the only case where
new ingredients, namely the four-parton initial-initial antennae 
are needed in unintegrated as well as in integrated forms.
On a longer term we are aiming to evaluate the whole set of 
integrated four-parton initial-initial antennae, those are 
universal building blocks for the subtraction terms 
for any hadronic process evaluated at NNLO. 

In a first step towards this aim, in this paper, we have focused 
on the crossings of two partons from a set of three 4-parton 
final-final antennae involving two quark flavours.
More precisely, the paper will be organised as follows:
In Section 2, we present the general formulae 
for the subtraction terms related to double real radiation 
for initial-initial configurations while the colour-connected case is treated 
explicitly in Section 3.
Section 4 establishes a list of all non-identical initial-initial 
four-parton antenna functions relevant to construct the subtraction term 
for the double real radiation off two initial-state partons.
In Section 5, the phase space mapping appropriate for initial-initial
configurations is presented. 

Finally, Section 6 contains our results for the integrated 
initial-initial antenna functions with two quark flavours 
and Section 7 our conclusions. Since the results are lengthy, 
we show only the leading pole terms in 
the manuscript, and attach the complete results as a {\tt Mathematica} file.

%********************

\section{Antenna subtraction for double-real radiation at NNLO 
in the initial-initial configuration} 

Antenna subtraction has been derived explicitly 
for final-final and initial-final configurations at NLO and NNLO 
in \cite{GehrmannDeRidder:2005cm} and in \cite{Daleo:2006xa,Daleo:2009yj} respectively.
For the initial-initial case, it has been derived at NLO 
in \cite{Daleo:2006xa} and is under construction at NNLO.

At NLO, the subtraction term is introduced to extract 
and cancel the infrared divergences present in the real contributions.
The general forms of the subtraction terms  
required at NLO in any of the three configurations 
(final-final, initial-final and initial-initial) have been given in 
\cite{GehrmannDeRidder:2005cm,Daleo:2006xa} and summarised in \cite{Joao}.
At this order, only tree-level three-parton antenna functions involving 
one unresolved parton are needed to build the subtraction terms. 
Those functions are usually denoted by 
$X^0_{ijk}$, $X^0_{i,jk}$, $X^0_{ik,j}$ in the three configurations.
Their definitions will be recalled in Section 4.

At NNLO, two contributions, double real and 
mixed real-virtual, require the introduction of a subtraction term.
For final-final and initial-final configurations both types of subtraction terms
have been constructed; at this order tree-level four-particle antennae 
involving two unresolved partons and one-loop three-parton antenna 
functions are needed respectively. Those functions have been derived   
and integrated over the corresponding factorised phase space  
in \cite{GehrmannDeRidder:2005cm} and \cite{Daleo:2009yj}.
For initial-initial configurations, 
the general structure of the unintegrated subtraction terms  
relevant for the double real contributions to $pp \to m \; \text{jets}$ 
is presented in \cite{Joao},
we will recall it in this section.

The double real radiation contribution to the $m$-jet cross section in $pp$ collisions reads
\begin{eqnarray}
\lefteqn{{\rm d}\hat\sigma^{R}_{NNLO}
= {\cal N}\,\sum_{m+2}{\rm d}\Phi_{m+2}( \psmomenta )
\frac{1}{S_{{m+2}}} }\nonumber \\ &\times&
|{\cal M}_{m+2}( \psmomentaini )|^{2}\;
J_{m}^{(m+2)}( \psmomenta )\;.
\label{eq:nnloreal}
\end{eqnarray}
In this equation,
$|{\cal M}_{m+2}( \psmomentaini )|^{2}$ stands for the colour-ordered 
$2 \rightarrow m+2$ matrix-element squared. 
The symmetry factor $S_{m+2}$ accounts for identical partons in the final state.
The normalisation factor ${\cal N}$ includes all QCD-independent factors 
as well as the dependence on the renormalised QCD coupling constant $\alpha_s$.
$\sum_{m+2}$ denotes the sum over all configurations with $m+2$ partons. 
 The initial state momenta are labelled as usual as $\pa$ and $\pb$ whereas the $m+2$ momenta in the final state are labeled $\fmomenta$. 
${\rm d}\Phi_{m+2}$ is the $2\to m+2$ particle phase space
\begin{eqnarray}
{\rm d}\Phi_{m+2}( \psmomenta )= \dmom{1}
\hdots 
\dmom{m+2} (2\pi)^d
\delta^d( \pa + \pb - \fmom{1} -\hdots- \fmom{m+2})
\end{eqnarray}
where we have introduced the abbreviation $ \dmom{1} = ( \d^{d-1} \fmom{1} )/( 2 E_1 (2 \pi)^{d-1} ) $. The jet function $J_{m}^{(m+2)}( \psmomenta )$ 
ensures that out of ($m+2$) final state partons, 
an observable with $m$ jets is built. 
The incoming parton momenta $\pa, \pb$ serve as reference directions to define transverse momenta and rapidities of the jets.

The double real radiation contribution given in eq.~\eqref{eq:nnloreal}
can become singular if either one or two final state partons 
are unresolved (soft or collinear).
Consequently, when constructing the corresponding subtraction term 
${\rm d}\hat\sigma^{S}_{NNLO}$ in eq.~\eqref{eq:sigNNLO}, which 
shall correctly subtract all those 
single and double unresolved singularities 
we must distinguish the following 
configurations according to the colour connection of the unresolved partons:
\begin{itemize}
\item[(a)] One unresolved parton but the experimental observable selects only
$m$ jets.
\item[(b)] Two colour-connected unresolved partons (colour-connected).
\item[(c)] Two unresolved partons that are not colour connected but share 
a common radiator (almost colour-unconnected).
\item[(d)] Two unresolved partons that are well separated from each other 
in the colour 
chain (colour-unconnected).
\item[(e)] Large angle soft gluon radiation. 
\end{itemize}

This separation among subtraction contributions according to colour connection
is valid in final-final, initial-final or initial-initial configurations 
and in any of those the subtraction formulae have a characteristic structure
in terms of required antenna functions.
This antenna  structure has been derived for the final-final and 
initial-final cases in \cite{GehrmannDeRidder:2005cm,Daleo:2009yj} 
and in \cite{Joao} for the initial-initial case. 

In here, we focus only on the initial-initial case 
with the kinematical situation where two unresolved partons are 
colour-connected to the two incoming partons. This is the only case where 
new ingredients are needed, namely the initial-initial four-parton 
antenna functions denoted by $X_{il,jk}$.
The four-particle initial-initial antenna functions will be defined 
explicitly in Section 4.

\section{Subtraction terms for two colour-connected
unresolved partons in the initial-initial configuration} 

When two unresolved partons $j$ and $k$ are adjacent and colour-connected to two initial-state partons,  
the subtraction term related to the double real contribution 
${\rm d}\hat\sigma^{R}_{NNLO}$ given in eq.~(\ref{eq:nnloreal}) reads :
\begin{eqnarray}
\lefteqn{{\rm d}\hat\sigma_{NNLO}^{S,b,(ii)}
=  {\cal N}\,\sum_{m+2}{\rm d}\Phi_{m+2}( \psmomenta )
\frac{1}{S_{m+2}} }\nonumber \\
&\times& \,\sum_{il} \sum_{jk}\;\left( X^0_{il,jk}
- X^0_{l,jk} X^0_{iL,K} - X^0_{i,kj} X^0_{Il,J} \right)\nonumber \\
&\times&
|{\cal M}_{m}( \reduced{ \imom{} }_{\hat{I} }, \reduced{ \imom{} }_{ \hat{L} }; \fmomentared )|^2\,
J_{m}^{(m)}( \fmomentared ; \reduced{ \imom{} }_{\hat{I} }, \reduced{ \imom{} }_{\hat{L} } )\;,
\label{eq:sub2bii}
\end{eqnarray}
where the sum runs over all colour-adjacent pairs $j,k$ and implies that 
the hard momenta $i,l$ are chosen accordingly.
 $X^0_{il,jk}$ denotes a four-particle tree-level initial-initial
antenna function defined explicitly in Section 4. 
By construction those contain all colour connected double unresolved limits 
of the $2 \rightarrow m+2$ parton matrix element associated with partons $j$ and $k$ 
being unresolved between radiators $i$ and $l$. 

However this antenna can also become singular in \textit{single} 
unresolved limits associated with $j$ or $k$ only where 
it does not coincide with limits of the matrix element. 
To ensure that this subtraction term is only active in 
the double unresolved limits of the real matrix elements squared 
we remove these single unresolved limits of the four-particle antennae.
Those limits are products of two tree-level three-particle antennae, namely 
products of an initial-final antenna $X_{a,bc}$ and an initial-initial antenna
$X_{Ad,C}$.  In these antennae we have replaced the original hard radiators 
with new particles, $\hat{I}$ and $\hat{L}$. 
When both radiators are in the initial state as it is the case here,
$p_{\hat{I}}= x_i p_i$, $p_{\hat{L}} = x_l p_l$.
The product of these three-particle antenna functions in 
${\rm d}\hat\sigma_{NNLO}^{S,b,(ii)}$ then subtracts the single unresolved 
limits of the associated four-particle antenna. 

The $2 \rightarrow m$ matrix element 
$|{\cal M}_{m}( \reduced{ \imom{} }_{ \hat{I} }, \reduced{ \imom{} }_{ \hat{L} }; \fmomentared )|^2$ 
\,is evaluated with new on-shell momenta which are Lorentz boosted as a result 
of the mapping required to ensure factorisation of matrix element 
and phase space. The NNLO initial-initial mappings have been discussed in 
\cite{Daleo:2006xa} and will be recalled in this paper in Section 5.  

\begin{figure}[t!]
\begin{center}
\includegraphics[width=0.8\textwidth]{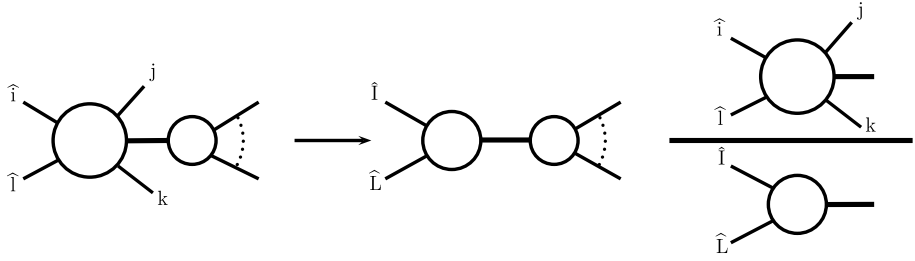}
\caption[Colour connected initial-initial antenna/phase space factorisation]{Illustration of NNLO antenna factorisation representing the
factorisation of both the squared matrix elements and the $(m+2)$-particle 
phase
space when the unresolved particles $j$ and $k$ are colour connected between two initial state radiators $\hat{i}$ and $\hat{l}$.}
\label{fig:biifact}
\end{center}
\end{figure}

Using a factorised form for both matrix element and phase space
as shown in figure \ref{fig:biifact},
one is able to obtain an integrated form of the subtraction term with 
$m$ partons in the final state 
which can be combined with the virtual two-loop contributions having also 
$m$ final state partons as defined in eq.~\eqref{eq:sigNNLO}.  

More explicitly,
the factorisation of the phase space with $m+2$ final state particles 
denoted by $\fmomenta$, reads, 
\begin{eqnarray}
\label{eq:psii4}
{\rm d}\Phi_{m+2}( \psmomenta )&=&
{\rm d}\Phi_{m}( \fmomentared ; \xa \pa, \xb \pb )
\nonumber\\
&&\times\delta( \xa - \xahat )\,\delta( \xb - \xbhat ) \dmom{j} \dmom{k} \d \xa \d \xb.
\end{eqnarray}
%with $\pa$ and $\pb$ being the initial state momenta.

Using this factorised form of the phase space we can rewrite 
the integrated (colour-connected) subtraction term involving 
only the four-parton antennae in the form,
\begin{eqnarray}
&&|{\cal M}_{m}|^2\,
J_{m}^{(m)}\; 
{\rm d}\Phi_{m}
\int \dmom{j} \dmom{k} \delta( \xa - \xahat )\,\delta( \xb - \xbhat ) \;X^0_{il,jk} \d \xa \d \xb.
\end{eqnarray}

Moreover, the integrated antennae are defined 
as the antenna functions integrated over the 
antenna phase space as defined in eq.~\eqref{eq:psii4} 
including a normalisation factor to account for powers of the 
QCD coupling constant by,
 \begin{eqnarray}
\label{eq:x4intii}
&&{\cal X}^0_{il,jk}(x_i,x_l, \ep)=\frac{1}{[C(\epsilon)]^2}\int \dmom{j} \dmom{k} \;x_i\;x_l\; \delta(x_i-\hat{x}_i)\,\delta(x_l-\hat{x}_l)\,X^0_{il,jk},\,
\end{eqnarray}

where $C(\epsilon)$ is given by,
\begin{equation}
C(\epsilon)=(4\pi)^{\epsilon}\frac{e^{-\epsilon\gamma}}{8\pi^2}.\label{eq:ceps}
\end{equation}

These integrations are performed analytically in $d$ dimensions to make 
the infrared singularities explicit. 
The integrated initial-initial antennae are presently unknown 
and the calculation of a sub-set of those involving two quark flavours 
is the subject of Section 6 and constitutes our main result in this paper. 
A list of all non-identical initial-initial four-particle antennae 
will be presented in Section 4.

\section{Initial-initial antenna functions}
In this section we shall recall how the antenna functions 
are defined in any of the configurations and how they enter in 
the construction of the subtraction terms. We also present a list of all 
non-identical four-parton initial-initial antenna functions.  
\subsection{Definitions}
In the antenna formalism, in any of the three configurations (final-final, initial-final or initial-initial) the subtraction terms are constructed 
from products of antenna functions with reduced matrix 
elements (with fewer final state 
partons than the original matrix element). The integrated subtraction
terms are obtained after an integration 
over a phase space which is factorised into an antenna phase space 
(involving all unresolved partons and the two radiators of the antenna) 
multiplied with a reduced phase space (where the momenta of radiators 
and unresolved radiation are replaced by two redefined momenta). 
These redefined momenta can be in the initial state,
if the corresponding radiator momenta were in the initial state 
as we saw in Section 3 in eq.~\eqref{eq:sub2bii} 
for the subtraction term $\sigma_{NNLO}^{S,b,(ii)}$.

An antenna function is 
determined by the external states it contains and the pair of hard
partons it collapses to in the unresolved limits. 
In general we denote the antenna function as $X$.
For antennae that collapse onto a hard quark-antiquark pair, 
$X = A$ for $qg \bar q$.  Similarly, for a quark-gluon antenna, we have 
$X = D$ for $qg g$ and $X=E$ for $qq^\prime
\bar q^\prime$ final states.  
Finally, we characterise the gluon-gluon antennae as $X=F$ for $ggg$,  
$X=G$ for $gq\bar q$ final states.

At NLO, we only need to consider tree level three-particle antennae 
involving only one unresolved parton. 
At NNLO we will need four-particle antennae involving 
two unresolved partons and one-loop three-particle antennae.

In all cases the antenna functions are derived from physical matrix 
elements associated to the decay of a colourless particle into partons: 
the quark-antiquark antenna functions are derived from 
$\gamma^* \to q\bar q~+$~(partons)~\cite{GehrmannDeRidder:2004tv}, 
the quark-gluon antenna functions from $\tilde\chi \to \tilde g~+$~(partons)
~\cite{GehrmannDeRidder:2005hi} and 
the gluon-gluon antenna functions from $H\to$~(partons)
~\cite{GehrmannDeRidder:2005aw}. The  tree-level antenna functions are obtained by normalising the three- and four-parton tree-level 
colour sub-amplitudes squared to that of the basic two-parton process:
The final-final three- and four-particle antennae are respectively defined by:
\begin{eqnarray}
X_{ijk}^0 = S_{ijk,IK}\, \frac{|{\cal M}^0_{ijk}|^2}{|{\cal M}^0_{IK}|^2}\;,\nonumber\\
X_{ijkl}^0 = S_{ijkl,IL}\, \frac{|{\cal M}^0_{ijkl}|^2}{|{\cal M}^0_{IL}|^2}\;.
\end{eqnarray}
where $S$ denotes the symmetry factor associated with the antenna, which accounts
both for potential identical particle symmetries and for the presence 
of more than one antenna in the basic two-parton process.
It is chosen such that the antenna function reproduces the unresolved 
limits of a matrix element with identified particles.

The initial-final tree-level three- and four-parton antennae denoted by $X_{i,jk}^0$ and  
$X_{i,jkl}^0$ are in principle obtained by crossing one-parton to the initial state starting 
from the corresponding final-final antennae. 
However this crossing might be ambiguous as was first noticed in \cite{Daleo:2006xa} 
for the quark-gluon type antenna. Depending on the unresolved limit considered, 
the pair of hard partons it collapses to may be different.

The initial-initial tree-level three- and four-parton antennae denoted 
by $X_{ik,l}^0$ and $X_{il,jk}^0$ are obtained by crossing two partons 
to the initial state, starting from the final-final antennae. 
This crossing procedure, unlike in the initial-final case, 
is free of ambiguity as the pair of hard partons the initial-initial 
antenna collapses to is always uniquely defined. 

More explicitly, 
the three-parton initial-initial antenna function $X^0_{ik,j}$ is defined as
\begin{equation}
	X^0_{ik,j} = S_{ijk,IK} \frac{ \abs{ \M \left( \cross{i}, j, \cross{k} \right) }^2 }{ \abs{ \M \left( \cross{I}, \cross{K} \right) }^2 } 
\end{equation}
where $\cross{i}$ denotes that particle $i$ is crossed to the initial state.
Therefore we can see that the initial-initial antenna is connected to the final-final antenna (where all coloured particles are outgoing) by
\begin{equation}
	X^0_{ik,j} = (-1)^{\Delta_F} X^0_3 \left( - p_i, p_j, -p_k \right)
\end{equation}
with $\Delta_F$ the difference in the number of fermion crossings between 
the three-particle and the two-particle subamplitude.

The tree-level four-particle initial-initial antenna $X^0_{il,jk}$ is defined as
\begin{equation}
\label{eq:antdef}
	X^0_{il, jk} = S_{ijkl,IL} \frac{\abs{ \M( \cross{i}, j, k, \cross{l} ) }^2 }{\abs{ \M( \cross{I}, \cross{L} ) }^2 }
\end{equation}
and the relation to the final-final antenna is
\begin{equation}
	X^0_{il, jk} = (-1)^{\Delta_F} X^0_4 \left( - p_i, p_j, p_k, -p_l \right).
\end{equation}

The four-particle tree-level antenna functions are not determined 
by the species of the particles alone but also by the colour-connection.
We distinguish leading-colour antennae, denoted by letters without tilde, 
where the particles are colour-connected in the order they are listed 
and subleading colour antennae, denoted by letters with tilde, where the gluons 
are photon-like. This notation has been used in 
~\cite{GehrmannDeRidder:2005cm,Daleo:2009yj}
and we will further use it in this section in the tables below.
 
The unresolved limits of the initial-initial antennae can be obtained from 
those of the final-final antennae by crossing. 
The crossing of the triple-collinear splitting functions 
is explained in \cite{deFlorian:2001zd}. Those limits will 
be reported elsewhere.

\subsection{Catalogue of tree-level four-particle initial-initial antenna functions}
\label{sec:catalogue}

Any two particles of a four-particle final-final antenna can be crossed 
to the initial state to obtain an initial-initial antenna; 
therefore one final-final four-particle antenna gives rise 
to six initial-initial antennae. 
Due to symmetries, at most four of these initial-initial antennae are different. 
The independent crossings are listed below. 
To make the colour connection clear, in this list we write out the arguments 
of the antennae explicitly, i.e we write $X^0_4 \left( \cross{i}, j, \cross{k}, l \right)$
(where $\cross{ i }$ denotes an incoming particle) instead of $X^0_{ik,jl}$.  
Some initial-initial antennae are free of singular limits. 
These finite antennae are not needed for the construction of subtraction terms, 
but their integrated form could be needed for cross-checks. 
They are marked with  $\limfreemark \limfreemark $ below.\\
\begin{center}
\begin{tabular*}{0.9 \textwidth}{l @{$\qquad$} l}
\toprule
\multicolumn{2}{l}{quark-antiquark antennae}\\
\midrule
$A^0_4$	&
$A^0_4 \big( \cross{ \q },	\cross{ \gluon }, 	\gluon,			\qb		 	\big)$,
$A^0_4 \big( \cross{ \q },	\gluon,			\cross{\gluon},		\qb			\big)$,
$A^0_4 \big( \cross{ \q },	\gluon,			\gluon,			\cross{\qb}	\big)$,
$A^0_4 \big( \q,			\cross{\gluon},		\cross{\gluon},		\qb			\big)$
\\[3pt]
$\widetilde{A}^0_4$	&
$\widetilde{A}^0_4 \big( \cross{ \q },	\cross{ \gluon }, 	\gluon,			\qb		 	\big)$,
$\widetilde{A}^0_4 \big( \cross{ \q },	\gluon,			\gluon,			\cross{\qb}	\big)$,
$\widetilde{A}^0_4 \big( \q,		\cross{\gluon},		\cross{\gluon},		\qb			\big)$
\\[3pt]
$B^0_4$	&
$B^0_4 \big( \cross{ \q },	\cross{ \qp }, 	\qpb,			\qb 			\big)$,
$B^0_4 \big( \cross{ \q },	\qp, 			\qpb,			\cross{ \qb } 	\big)$,
$B^0_4 \big( \q,		\cross{ \qp }, 	\cross{ \qpb },	\qb
\big)^{\limfreemark \limfreemark}$
\\[3pt]
$C^0_4$	&
$C^0_4 \big( \cross{ \q },	\cross{ \qb },	\q,			\qb	 		\big)$,
$C^0_4 \big( \cross{ \q },	\qb, 			\cross{ \q},		\qb	 		\big)$,
$C^0_4 \big( \q,		\cross{ \qb },	\cross{ \q },	\qb
\big)^{\limfreemark \limfreemark}$,
$C^0_4 \big( \q,		\qb,			\cross{ \q },
\cross{ \qb }	\big)^{\limfreemark \limfreemark}$ \\
\bottomrule
\end{tabular*}
\end{center}
$\widetilde{A}^0_4 \left( \cross{ \q },	\gluon,			\cross{\gluon},		\qb			\right)$
is symmetric under the interchange of the two photon-like gluons.
The nonidentical-flavour antenna $B^0_4$ is separately symmetric under interchange of $\qp$ with $\qbp$ and of $\q$ with $\qb$.
\begin{center}
\begin{tabular*}{0.9 \textwidth} { l @{$\qquad$} l }
\toprule
\multicolumn{2}{l}{quark-gluon antennae}\\
\midrule
$D^0_4$	&
$D^0_4 \big( \cross{ \q },	\cross{ \gluon }, 	\gluon,			\gluon 			\big)$,
$D^0_4 \big( \cross{ \q },	\gluon,	 		\cross{ \gluon },		\gluon 			\big)$,
$D^0_4 \big( \q,		\cross{ \gluon },		\cross{ \gluon },		\gluon 			\big)$,
$D^0_4 \big( \q,		\cross{ \gluon },		\gluon,			\cross{ \gluon } 		\big)$
\\[3pt]
$E^0_4$	&
$E^0_4 \big( \cross{ \q },	\cross{ \qp },		\qbp,				\gluon			\big)$,
$E^0_4 \big( \cross{ \q },	\qp,				\qbp,				\cross{ g }			\big)$,
$E^0_4 \big( \q,		\cross{ \qp },	\cross{ \qbp },	\gluon			\big)$,
$E^0_4 \big( \q,		\cross{ \qp },	\qbp,			\cross{ g }		\big)$
\\[3pt]
$\widetilde{E}^0_4$	&
$\widetilde{E}^0_4 \big( \cross{ \q },	\cross{ \qp },		\qbp,				\gluon			\big)$,
$\widetilde{E}^0_4 \big( \cross{ \q },	\qp,				\qbp,				\cross{ g }			\big)$,
$\widetilde{E}^0_4 \big( \q,		\cross{ \qp },	\cross{ \qbp },	\gluon			\big)$,
$\widetilde{E}^0_4 \big( \q,		\cross{ \qp },	\qbp,			\cross{ g }		\big)$
\\[3pt]
\bottomrule
\end{tabular*}
\end{center}
Due to the cyclic colour connection, $D^0_4$ is symmetric under interchange of the second and fourth gluon.

\begin{center}
\begin{tabular*}{0.9 \textwidth} { l @{$\qquad$} l }
\toprule
\multicolumn{2}{l}{gluon-gluon antennae}\\
\midrule
$F^0_4$	&
$F^0_4 \big( \cross{ \gluon },	\cross{ \gluon }, 	\gluon,			\gluon 			\big)$,
$F^0_4 \big( \cross{ \gluon },	\gluon,	 		\cross{ \gluon },		\gluon 			\big)$
\\[3pt]
$G^0_4$	&
$G^0_4 \big( \cross{ \gluon },			\cross{ \q },		\qb,			\gluon			\big)$,
$G^0_4 \big( \cross{ \gluon },			\q,				\cross{ \qb },	\gluon			\big)$,
$G^0_4 \big( \cross{ \gluon },			\q,				\qb,			\cross{ \gluon }		\big)$,
$G^0_4 \big( \gluon,				\cross{ \q },		\cross{ \qb },		\gluon			\big) $
\\[3pt]
$\widetilde{G}^0_4$	&
$\widetilde{G}^0_4 \big( \cross{ \gluon },			\cross{ \q },		\qb,			\gluon			\big)$,
$\widetilde{G}^0_4 \big( \cross{ \gluon },			\q,				\qb,			\cross{\gluon}		\big)$,
$\widetilde{G}^0_4 \big( \gluon,				\cross{ \q },		\cross{ \qb },	\gluon			\big) $
\\[3pt]
$H^0_4$ &
$H^0_4 \big(		\cross{ \q },		\cross{ \qb },		\qp,			\qbp			\big)$,
$H^0_4 \big(		\cross{ \q },		\qb,				\cross{ \qp },	\qbp			\big)$
\\[3pt]
\bottomrule
\end{tabular*}
\end{center}
$F^0_4$ is symmetric under cyclic interchange of its arguments.
$\widetilde{G}^0_4$ is symmetric under the interchange of the two gluons as well 
as under the interchange of $\q$ with $\qb$.
$H^0_4$ has three symmetries, $\q \leftrightarrow \qb$, $\qp \leftrightarrow \qbp$ 
and the flavour renaming $\q \leftrightarrow \qp$, $\qb \leftrightarrow \qbp$.

\section{Phase space factorisation and mappings}
\label{sec:PSfact}

The construction of subtraction terms requires a mapping from the original set
of momenta onto a reduced set. The mapping interpolates between the different
soft and collinear limits which the subtraction term regulates.
An appropriate mapping for the initial-initial case, both for single
and double unresolved configurations, has been discussed
in~\cite{Daleo:2006xa}. By requiring momentum conservation and phase space
factorisation, the phase space mapping is strongly constrained.
The remapping of initial state momenta can only be a rescaling, since
any transversal component would spoil the phase space factorisation.
For two unresolved partons $j$ and $k$, a complete factorisation of the phase
space into a convolution of an $m$-particle phase space
depending on redefined momenta only and the phase space of the unresolved 
partons $j$ and $k$ can be achieved with a Lorentz boost.
This boost maps the momentum $q \;=\; \pa + \pb - \fmom{j} - \fmom{k} $\,, with $q^2>0$\,
and $\pa, \pb$ being the momenta of the hard emitters,
into the momentum $\tilde{q} \,=\, \xa \pa + \xb \pb $\,, where $\xa$ and $\xb$
are fixed in terms of the invariants as follows:
\begin{eqnarray}
\xa &=&\left(\frac{s_{12}-s_{j2}-s_{k2}}{s_{12}}
\;    \frac{s_{12}-s_{1j}-s_{1k}-s_{j2}-s_{k2}+s_{jk}}{s_{12}-s_{1j}-s_{1k}}
     \right)^{\frac{1}{2}}\,,\nonumber\\
\xb &=&\left(\frac{s_{12}-s_{1j}-s_{1k}}{s_{12}}
\;    \frac{s_{12}-s_{1j}-s_{1k}-s_{j2}-s_{k2}+s_{jk}}{s_{12}-s_{j2}-s_{k2}}
     \right)^{\frac{1}{2}}\,.
\end{eqnarray}
These two definitions guarantee the overall momentum conservation in the
mapped momenta and the correct soft and collinear behaviours\,.
The two momentum fractions $\xa$ and $\xb$ satisfy the following limits in
double unresolved configurations:
\begin{enumerate}
\item $j$ and $k$ soft: $\xa\rightarrow 1$, $\xb\rightarrow 1$,
\item $j$ soft and $k_{k}=z_1p_1$: $\xa\rightarrow 1-z_1$, $\xb\rightarrow 1$,
\item $k_{j}=z_1p_1$ and $k_{k}=z_2p_2$:
  $\xa\rightarrow 1-z_1$, $\xb\rightarrow 1-z_2$,
\item $k_{j}=z_1p_1$, $\fmom{k} = z_2 p_1$: $\xa\rightarrow 1-z_1-z_2$,
  $\xb\rightarrow 1$,
\end{enumerate}
and all the limits obtained from the ones above by the exchange of $p_1$ with
$p_2$ and of $k_j$ with $k_k$.
The construction of NNLO antenna subtraction terms also requires that
all single unresolved limits of the four-parton initial-initial antenna
functions $X_{il,jk}$, with radiators $i$ and $l$,
have to be subtracted, such that the resulting subtraction term is active only
in its double unresolved limits. A systematic subtraction of these single
unresolved limits by products of two three-parton antenna functions can be
performed only if the NNLO phase space mapping turns into an NLO phase space
mapping in its single unresolved limits. A detailed discussion of
the corresponding translation between these 
two momentum mappings can be found in~\cite{Daleo:2006xa}.
\\

The factorisation of the $(m+2)$-parton phase space into 
an $m$-parton phase space and an antenna phase space involving the unresolved
partons $j$ and $k$ given in eq. \eqref{eq:psii4} can equivalently
be written as
\begin{equation}
\begin{split}
\d\Phi_{m+2}( \psmomenta )&=
\d\Phi_{m}(\fmomentared;\xa \pa,\xb \pb)
\\
&\quad \times \; {\cal J} \;\delta(q^2-\xa\,\xb\,s_{12})\,
\delta(2\,(\xb \pb-\xa \pa ). q)\,
\\
& \quad \times \;[\d k_j]\;[\d k_k]\;\d \xa\;\d \xb\,,
\label{PS}
\end{split}
\end{equation}
where ${\cal J}$ is the Jacobian factor defined by
\[
{\cal J}=s_{12}\,\left(\xa(s_{12}-s_{1j}-s_{1k}) + \xb
(s_{12}-s_{2j}-s_{2k})\right)\,.
\]

This phase space parametrization can also be used to give an 
equivalent definition of the integrated initial-initial antennae first given in 
eq.~(\ref{eq:x4intii}).
Those are given as,  
\begin{equation}
\label{eq:intant}
{\cal X}^0_{il, jk}(x_1,x_2,\ep)=\frac{1}{[C(\epsilon)]^2}\int [{\rm d} k_j][{\rm d} k_k] \; {\cal J}\; x_1\; x_2\; \delta(C_1)\,\delta(C_2)\,X^0_{il,jk}\,,
\end{equation}
where
\begin{eqnarray}
\label{constarints}
  \nonumber
  C_1 &=& q^2 - \xa \, \xb  \, s_{12}\,, \\
  C_2 &=& 2 (\xb \pb - \xa \pa) . q\,.
\end{eqnarray}

We shall use this definition of the integrated antennae $
{\cal X}^0_{il, jk}(x_1,x_2,\ep)$ to compute them in Section 6.
\section{Integration of the four-parton initial-initial antennae}
\label{sec:calc}

In the first part of this section we describe how the initial-initial 
four-parton antennae are integrated over the antenna phase space 
while in the second part of this section we restrict ourselves 
to the evaluation of integrated antennae involving two quark flavours. 

\subsection{Calculational method}
The initial-initial antenna functions have the scattering kinematics
\begin{displaymath}
\pa + \pb \to \fmom{j} + \fmom{k} + q \;,
\end{displaymath}
where $q$ is the momentum of the outgoing colourless particle.
The momenta satisfy:
\begin{displaymath}
\pa^2=\pb^2 =0, \;\;  \fmom{j}^2= \fmom{k}^2 =0, \;\;
q^2 = \tilde{q}^2 = x_1 \,x_2 \,s_{12}\,.
\end{displaymath}
The four-parton initial-initial antennae defined in Section 4  
need to be integrated over the phase space of the unresolved partons $j$ 
and $k$. This integration yields a result which depends only on $s_{12}$, 
$\xa$ and $\xb$. From dimensional
counting, one can immediately conclude that the dependence on $s_{12}$
is only multiplicative, according to the mass dimension of the integral.
\\
The initial-initial antenna phase space integrals are derived 
from squared matrix elements and can be represented 
by forward scattering diagrams as in the following figure:
%
%\begin{figure}[h!]
\begin{center}
   \includegraphics[width=0.50\textwidth]{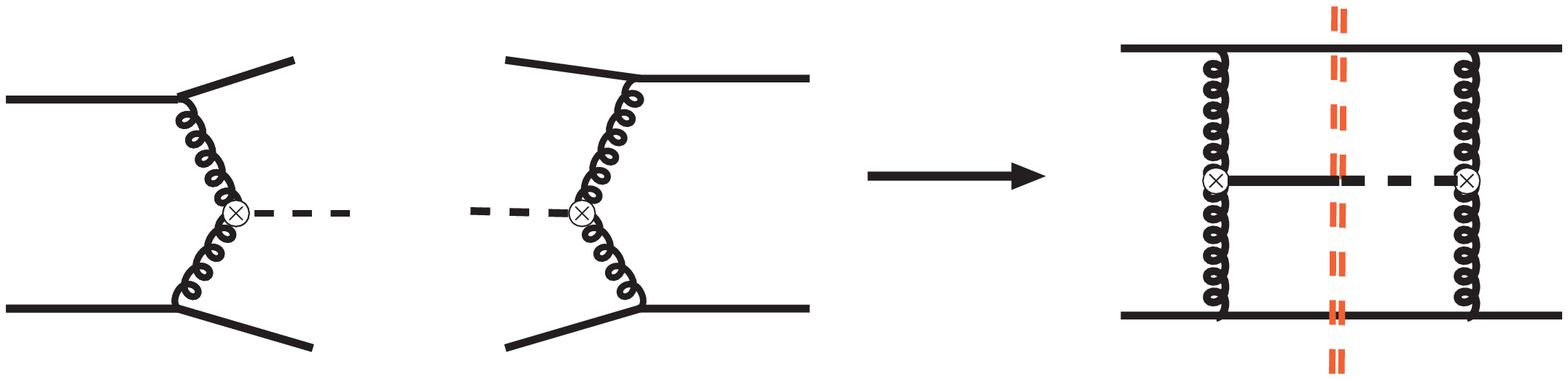}
   \label{fig:cutantennae}
\end{center}
%\end{figure}
%
The two delta functions in eq.~(\ref{PS}) can be represented
as mass-shell conditions of fake particles and are shown in the previous
picture as a thick solid line (representing a massive particle with mass
$M^2= \xa \, \xb \,s_{12}$)
and a dashed line (representing the other constraint)\,.
This allows us to use the optical theorem
to transform the initial-initial antenna phase space integrals into cut
two-loop integrals and therefore use the methods developed for
multi-loop
calculations~\cite{Anastasiou:2002yz,Anastasiou:2003ds,Anastasiou:2003yy}\,.
Up to eight-propagator integrals with four cut propagators are generated
in this way\,.
Using the reduction techniques,
the calculation of the integrated antennae can be related
to the evaluation of a reduced set of master integrals\,. For the complete set
of non-identical initial-initial four-parton antennae tabulated in Section 4,
we find $32$ such integrals, obtained using 
integration-by-part (IBP,~\cite{chet})
and Lorentz invariance (LI,~\cite{gr}) identities, following the Laporta
algorithm~\cite{laporta}. A private implementation as well as a public 
one~\cite{Smirnov:2008iw} have been used.
\\
For those four-parton antenna functions 
there are $13$ different propagators, including the four that are cut 
in the phase space integration ($D_{j}$, $D_{k}$, $D_{jk12}$, $D_{jk123}$):
\begin{eqnarray}
D_{j1} &=& (p_1-k_j)^2 \;,\nonumber \\
D_{k1} &=& (p_1-k_k)^2\;, \nonumber \\
D_{j2} &=& (p_2-k_j)^2\;, \nonumber \\
D_{k2} &=& (p_2-k_k)^2\;, \nonumber \\
D_{jk} &=& (k_j+k_k)^2\;, \nonumber \\
D_{jk1} &=& (p_1-k_j-k_k)^2\;, \nonumber \\
D_{jk2} &=& (p_2-k_j-k_k)^2\;, \nonumber \\
D_{j12} &=& (p_1+p_2-k_j)^2\;, \nonumber \\
D_{k12} &=& (p_1+p_2-k_k)^2\;, \nonumber \\
D_{j} &=& k_j^2\;, \nonumber \\
D_{k} &=& k_k^2\;, \nonumber \\
D_{jk12} &=& (p_1+p_2-k_j-k_k)^2 - x_1 x_2 s_{12}\;, \nonumber\\
D_{jk123} &=& (p_3 + p_1 + p_2 - k_j-k_k)^2 \;,
\end{eqnarray}
where $p_3 = x_2 \,p_2 -x_1 \,p_1$. To perform the reduction to master
integrals, we drop any integral where $D_{j}$, $D_{k}$, $D_{jk12}$,
$D_{jk123}$ are not in the denominator and impose momentum conservation.
The integrands of the $32$ master integrals found can all be written as 
rational polynomials of the denominators above.

These master integrals are calculated using either the
method of differential equations or by a direct
evaluation of the phase space integrals in terms of 
hypergeometric functions. The simplest master integral is 
a two loop bubble with all the internal lines cut, 
it is obtained from eq.~\eqref{eq:intant} by replacing the jacobian 
$\mathcal{J}$ and the antenna $X^0_{il,jk}$ with unity:\\
\begin{eqnarray}
\label{masterPR1}
&&\PR1 \;\;= \,I_2(x_1,x_2) \,= \,\int d^d q \,d^dk_j \,d^d k_k\,
                       \delta^d\left(p_1+p_2-q-k_j-k_k\right) \;
\times \nonumber
\\&& \hspace{4cm}
        \delta^+\left(k_j^2\right) \,
        \delta^+\left(k_k^2\right)\,\delta^+\left(q^2-M^2\right)
        \delta(2\,\left(x_2p_2-x_1p_1).q\right) \,.
\end{eqnarray}
This integral can actually be expressed as a hypergeometric function:
\begin{equation}
\begin{split}
	I_2 	&= 	s_{12}^{- 2 \ep} \frac{ ( 4 \pi )^{-4+2 \ep} }{ \Gamma ( 2 - 2 \ep) }
				\frac{ \Gamma( 1 - \ep )^2 }{ \Gamma( 2 - 2 \ep ) }
				x_1^{-\ep} (1-x_1)^{1-2 \ep} (1+x_1)^{- \ep} \\
		& \quad	x_2^{-\ep} (1-x_2)^{1-2 \ep} (1+x_2)^{- \ep }
				( x_1 + x_2 )^{-1+2 \ep } \\
		& \quad	_2F_1 \left( \ep, 1 - \ep, 2 - 2 \ep, \frac{ (1 - x_1)(1-x_2) }{ (1+x_1) (1+x_2) } \right),
\end{split}	
\end{equation}
it has been checked against the master integral $I[0]$ appearing in the 
calculation of the gauge-boson rapidity distribution at NNLO in~\cite{Anastasiou:2003ds}, 
the notation translates as $u = x_2/x_1$, $z = x_1 x_2$.\\
The set of master integrals which we denote by ${ I}_i(x_1,x_2,\ep)$ 
are functions of $x_1$, $x_2$ and $\ep$.  
We begin by factoring out the leading behavior of the master integrals
$I_i(x_1,x_2, \ep)$ in the limits $x_1 \to 1$
and $x_2 \to 1$, keeping the exact $\ep$-dependence:
\begin{equation}
{ I}_i(x_1,x_2,\ep) = (1-x_1)^{m_1-2 \,\ep} \;(1-x_2)^{m_2 - 2 \,\ep}\;
{\cal F}_i(x_1,x_2,\ep).
\end{equation}
The integers $m_1,m_2$ are characteristic to each master integral.
The functions ${\cal F}_i(x_1,x_2,\ep)$ are regular  at $x_1=1$,
at $x_2=1$, and at $x_1= x_2\, =1$
and can be calculated as Laurent series with, at most, second order
poles in $\ep$.

The integrated antennae given by ${\cal X}(x_1,x_2,\ep)$ are linear
combinations of these master integrals ${ I}_i(x_1,x_2,\ep)$, with
coefficients containing poles in $\ep$, as
well as in $(1-x_1)$ and $(1-x_2)$. After the masters have been inserted
into the integrated antennae, those take the form
%---------------
\begin{equation}
{\cal X}(x_1,x_2,\ep)
 =
(1-x_1)^{-1-2\,\ep}
(1-x_2)^{-1-2\,\ep}\;
{\cal R}(x_1,x_2,\ep).
\end{equation}
%---------------
where ${\cal R}(x_1,x_2,\ep)$ is regular at the boundaries $x_1 = 1$,
$x_2 = 1$, and at $x_1 = x_2 = 1$. The $\ep$-expansion of the
singular factors $(1-x_i)^{-1-2\ep}$ is done in the form
of distributions:
%---------------
\begin{equation}
\label{plusDist}
(1-x_i)^{-1-2\ep}\,=\, -\frac{1}{2\ep} \, \delta(1-x_i) + \sum_{n}
\frac{(-2\ep)^{n}}{n!}\mathcal{D}_{n}(x_i)\,,
\end{equation}
with
\begin{equation}
\mathcal{D}_{n}(x_i)=\l(\frac{\ln^{n}\l(1-x_i\r)}{1-x_i}\r)_{+}\,.
\end{equation}
%---------------

To evaluate the integrated antennae,
we decompose the phase space into four regions depending on the values 
of $x_1$ and $x_2$. Those regions are given by:
\begin{itemize}
\item $x_1\,\neq\,1$, \,$x_2\,\neq\,1$, which we refer to as the hard region
\item $x_1\,=\,1$,\, $x_2\,\neq\,1$, and  $x_1\,\neq\,1$,\, $x_2\,=\,1$,
referred to as collinear regions
\item $x_1\,=\,1$,\, $x_2\,=\,1$, which we denote the soft region\,.
\end{itemize}
%
%In each of these regions, the calculation of the integrated antennae as well
%as of the masters integrals involved is performed differently:
In the hard region ($x_1 \neq 1,\, x_2 \neq 1$), 
harmonic polylogarithms of weight two appear 
in the ${\cal O}(\ep^0)$ term of ${\cal R}$. Therefore,
the $\ep$-expansion of the master integrals in the hard region is needed 
at least up to the order at which terms of 
transcendentality\footnote{We define the transcendentality of a product as the sum of the 
transcendentalities of its factors, GHPL's and HPL's have transcendentality 
equal to their weight, $\log 2$ has transcendentality one, $\zeta(n)$ has 
transcendentality $n$.} 
two appear for the first time generally.

In the collinear regions ($x_1 = 1$ or $x_2 = 1$), 
since the expansion in distributions~(\ref{plusDist}) generates  
additional $1/\ep$ factors, the function ${\cal R}$ is required 
up to ${\cal O}(\ep)$ where harmonic polylogarithms of weight 3 appear.
The masters evaluated in the collinear region need therefore to be expanded 
at least up to the order at which terms of transcendentality 
3 appear generally. 

Finally, in the soft region ($x_1=x_2=1$),
since the expansion of the distributions~(\ref{plusDist}) generates
additional $1/\ep^2$ coefficients, the function ${\cal R}$ is required 
up to ${\cal O}(\ep^2)$ where polylogarithms of weight 4 appear.
The masters evaluated in the soft region need therefore to be expanded 
up to transcendentality 4 at least. 

\subsection{Integrated antennae with two quark flavours}

In a first step towards the calculation
of all integrated four-parton initial-initial antenna functions for
the double real radiation case,
in this paper, we focus on all the crossings of two partons from the following
four-parton final-final antennae: $B_4^0(q,q',\bar{q}',\bar{q})$,
$\tilde{E}_4^0(q,q',\bar{q}',g)$ and  $H_4^0(q,\bar{q},q',\bar{q}')$ defined
in~\cite{GehrmannDeRidder:2005cm}\,.

\begin{figure}[t!]
 \begin{center}
   \includegraphics[width=0.70\textwidth]{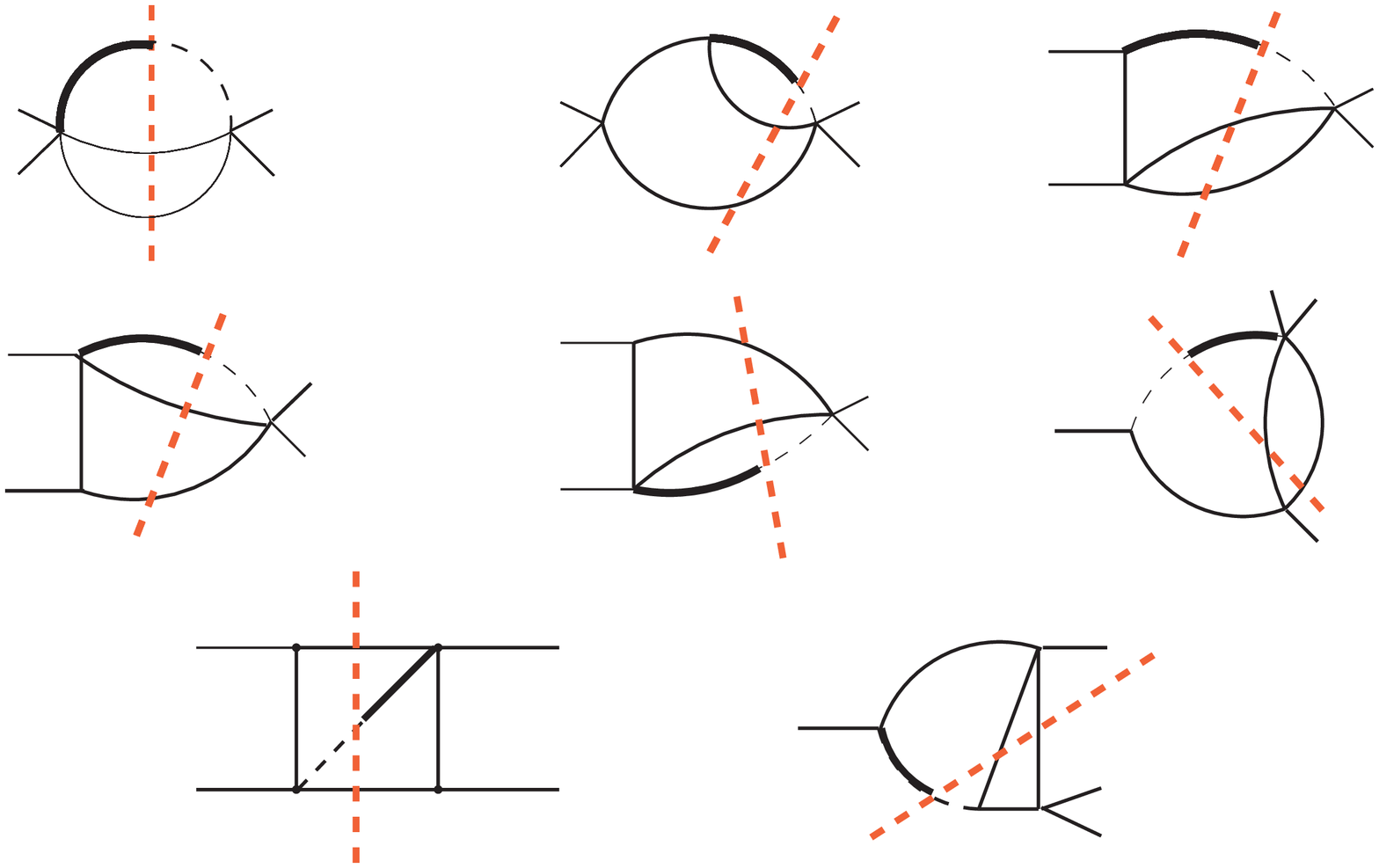}
   \caption{\textsf{Master integrals for the phase space integration
   of the tree-level initial-initial $\mathcal{B}_4^0,\,
   \mathcal{H}_4^0$ and $\mathcal{\tilde{E}}_4^0$
   type antennae
   at NNLO\,. Thick solid and
   dashed lines refer to the conditions on the phase space integral
   implemented as auxiliary propagators\,.
   All the internal lines are massless except for the thick solid line\,.
   Only the integrals without numerators are shown in this picture\,. }}
   \label{fig:mastersHBEtnew}
 \end{center}
\end{figure}

We found that the reduction involving only those initial-initial antennae 
leads to $12$ master integrals. Those without numerators are shown in
Fig.~\ref{fig:mastersHBEtnew}\,.
We have performed the calculation of these integrated antennae with two choices
of master integral bases differing by four master integrals.
In the first basis, 
the definitions of the master integrals involved in the calculation 
are as follows:
\begin{eqnarray}
% {1, SS[kb, p1]},
I_{1} & = & \int \dmom{j} \dmom{k} \delta(C_1) \delta(C_2)
        ( \fmom{k} \cdot p_1 )\,,
\\\nonumber
% {2, 1},
I_{2} & = & \int \dmom{j} \dmom{k} \delta(C_1) \delta(C_2) \,,
\\\nonumber
% {3, Db12 SS[ka, p1] SS[kb - p1 - p2, p3]},
I_{3} & = & \int \dmom{j} \dmom{k} \delta(C_1) \delta(C_2)
      \frac{  - (\fmom{j} \cdot p_1) (\fmom{j} \cdot p_3)}{D_{k12}} \,,
\\\nonumber
% {4, Db12 SS[ka, p1]},
I_{4} & = & \int \dmom{j} \dmom{k} \delta(C_1) \delta(C_2)
      \frac{  (\fmom{j} \cdot p_1) }{D_{k12}}\,,
\\\nonumber
% {5, Db12 SS[kb - p1 - p2, p3]},
I_{5} & = & \int \dmom{j} \dmom{k} \delta(C_1) \delta(C_2)
      \frac{ - (\fmom{j} \cdot p_3) }{D_{k12}} \,,
\\\nonumber
% {6, Db12},
I_{6} & = & \int \dmom{j} \dmom{k} \delta(C_1) \delta(C_2)
      \frac{ 1 }{D_{k12}} \,,
\\\nonumber
% {7, Dab2},
I_{7} & = & \int \dmom{j} \dmom{k} \delta(C_1) \delta(C_2)
      \frac{ 1 }{D_{jk2}} \,,
\\\nonumber
% {8, Db2},
I_{8} & = & \int \dmom{j} \dmom{k} \delta(C_1) \delta(C_2)
      \frac{ 1 }{D_{k2}} \,,
\\\nonumber
% {9, Db1},
I_{9} & = & \int \dmom{j} \dmom{k} \delta(C_1) \delta(C_2)
      \frac{ 1 }{D_{k1}} \,,
\\\nonumber
% {10, Dab1},
I_{10} & = & \int \dmom{j} \dmom{k} \delta(C_1) \delta(C_2)
      \frac{ 1 }{D_{jk1}} \,,
\\\nonumber
% {14, Da12 Dab2},
I_{14} & = & \int \dmom{j} \dmom{k} \delta(C_1) \delta(C_2)
      \frac{ 1 }{D_{j12} D_{jk2}}\,,
\\\nonumber 
% {15, Da1 Db2}
I_{15} & = & \int \dmom{j} \dmom{k} \delta(C_1) \delta(C_2)
      \frac{ 1 }{D_{j1} D_{k2}} \,,
\end{eqnarray}

In the other choice of basis, the masters with scalar products in the numerator 
($I_{1},I_{3},I_{4},I_{5}$) are replaced by alternative 
master integrals:
\begin{equation}
\begin{split}
I_{1}'		&=	\int \dmom{j} \dmom{k} \delta(C_1) \delta(C_2) D_{k12} \\
I_3'		&=	\int \dmom{j} \dmom{k} \delta(C_1) \delta(C_2) \frac{D_{j12}}{D_{k12}} \\
I_4'		&=	\int \dmom{j} \dmom{k} \delta(C_1) \delta(C_2) \frac{D_{j2}}{D_{k12}} \\
I_5'		&=	\int \dmom{j} \dmom{k} \delta(C_1) \delta(C_2) \frac{1}{D_{jk} D_{k12}}.
\end{split}
\end{equation}

The relation to the first basis is:
\begin{eqnarray} 
I_1 &= &s_{12} \frac{1}{4} I_2 - \frac{x_2}{2 (x_1 + x_2) } I_1'\nonumber \\ 
I_3 &= &s_{12} x_2 \frac{2 x_1^2 + 9 x_1 x_2 + x_2^2 - 12 x_1^2 x_2^2 + 
 2 \epsilon (-2 x_1^2 - 7 x_1 x_2 - x_2^2 + 10 x_1^2 x_2^2)}{8 (-1+2 \epsilon) (x_1-x_2) (x_1+x_2)}I_2 \\ \nonumber
&& - s_{12} x_1 \left[  \frac{-x_1^2 + x_1 x_2 - 4 x_2^2 - 4 x_1^2 x_2^2 + 8 x_1 x_2^3}{8 (-1+2 \epsilon) (x_1-x_2) (x_1+x_2)} \right. \\ \nonumber 
&& \qquad \qquad + \left. 2 \epsilon \frac{x_1^2 - x_1 x_2 + 2 x_2^2 + 4 x_1^2 x_2^2 - 6 x_1 x_2^3}{8 (-1+2 \epsilon) (x_1-x_2) (x_1+x_2)} \right] I_3' \\ \nonumber
&& + s_{12} \frac{1}{8} ( x_1 + x_2 + 4 x_1 x_2^2 ) I_4' \\ \nonumber
&& - s_{12}^3 \epsilon \frac{(-1 + x_1) x_1 (1 + x_1) (-1 + x_2) x_2^2 (1 + x_2) (-1 + x_1 x_2)}{2 (-1+2 \epsilon) (x_1-x_2) (x_1+x_2)} I_5' \\ \nonumber
&& + s_{12}^2 x_1 \left[ \frac{-(x_1 - x_2) (x_1 + x_2) (1 + 2 x_2^2)}{8 (-1+2 \epsilon) (x_1-x_2) (x_1+x_2)} \right. \\ \nonumber
&& \qquad \qquad \left. + 2 \epsilon \frac{x_1^2 - 3 x_2^2 + 2 x_1^2 x_2^2 - 2 x_2^4 + 2 x_1^2 x_2^4}{8 (-1+2 \epsilon) (x_1-x_2) (x_1+x_2)} \right] I_6\nonumber \\ 
I_4 &= &- \frac{1}{2} I_3' + \frac{1}{2} I_4' + s_{12} \frac{1}{2} I_6\nonumber \\ 
I_5 &= &- \frac{x_1}{2} I_3' + \frac{x_1+x_2}{2} I_4' + s_{12} \frac{x_1}{2} I_6.\nonumber
\end{eqnarray} 

Before we proceed with the details of calculating the masters, we present
in Table~\ref{tab:summarySC} a summary of which regions
contribute to the crossings of the antenna functions
$B_4^0(q,q',\bar{q}',\bar{q})$,
$\tilde{E}_4^0(q,q',\bar{q}',g)$ and
$H_4^0(q,\bar{q},q',\bar{q}')$.
We then present results of the required masters in each of these regions. 
As explained in Section \ref{sec:catalogue}, not all of the six crossings of a 
final-final antenna are different. Labelling the final-final antenna functions
as $B_4^0(1q,3q',4\bar{q}',2\bar{q})$, $\tilde{E}_4^0(1q,2q',3\bar{q}',4g)$
and $H_4^0(1q,2\bar{q},3q',4\bar{q}')$, the identical crossings are
the following:
\begin{eqnarray}
% H12 = H34 ;
  &&  H_{12}  =  H_{34}
\nonumber \\
% H13 = H14 = H23 = H24 ;
  &&  H_{13}  =  H_{14} = H_{23} = H_{24}
\nonumber \\
% B13 = B23 = B14 = B24 ;
  &&  B_{13}  =  B_{14} = B_{23} = B_{24}
\nonumber \\
% Et12 = Et13 ;
  &&  \tilde{E}_{12} = \tilde{E}_{13}
\nonumber \\
% Et24 = Et34 ;
  &&  \tilde{E}_{24} = \tilde{E}_{34}
\end{eqnarray}
where
\begin{eqnarray}
H_{12} &=& H_{\bar{q}q,q'\bar{q}'}^0 \,,
\\
H_{13} &=& H_{\bar{q}\bar{q}',\bar{q}\bar{q}'}^0 \,,
\\
B_{12} &=& B_{\bar{q}q,q'\bar{q}'}^0 \,,
\\
B_{34} &=& B_{\bar{q}'q',q\bar{q}}^0 \,,
\\
B_{13} &=& B_{\bar{q}\bar{q}',\bar{q}'\bar{q}}^0 \,,
\\
\tilde{E}_{12} &=& \tilde{E}_{\bar{q}\bar{q}',\bar{q}'g}^0 \,,
\\
\tilde{E}_{14} &=& \tilde{E}_{\bar{q}g,q'\bar{q}}^0 \,,
\\
\tilde{E}_{23} &=& \tilde{E}_{\bar{q}'q',q'g}^0 \,,
\\
\tilde{E}_{24} &=&\tilde{E}_{\bar{q}'g,q\bar{q'}}^0 \,.
\end{eqnarray}
As mentioned in Section \ref{sec:catalogue}, $B_{34}$ is free of singular limits. It will therefore not be needed for the construction of subtraction terms and its integrated form is free of poles in $\ep$. It might be needed for checks of the integrated antennae, however, which is why it is included here.
\begin{table}[th!]
\begin{center}
\begin{tabular}{ | c || c || c || c || c ||}
\hline
Antenna&soft&collinear\,$x_1$=1&collinear $x_2$=1&hard
\\[.5em]
\hline\hline
$\mathcal{H}_{12}$&no&no&no&yes\\[.5em]
\hline
$\mathcal{H}_{13}$&no&no&no&yes\\[.5em]
\hline
$\mathcal{\tilde{E}}_{12}$&no& yes&no&yes
\\[.5em]
\hline
$\mathcal{\tilde{E}}_{14}$&no&yes&no&yes
\\[.5em]
\hline
$\mathcal{\tilde{E}}_{23}$&no&no&no&yes\\[.5em]
\hline
$\mathcal{\tilde{E}}_{24}$&no&no&no&yes\\[.5em]
\hline
$\mathcal{B}_{12}$&yes&yes&yes&yes
\\[.5em]
\hline
$\mathcal{B}_{13}$&no&yes&no&yes
\\[.5em]
\hline
$\mathcal{B}_{34}$&no&no&no&yes\\[.5em]
\hline
\end{tabular}
\caption{Summary of the regions contributing to each of the independent
crossings of the three antennae: $\mathcal{B}_4^0,\;\mathcal{\tilde{E}}_4^0,\;
\mathcal{H}_4^0$.}
\label{tab:summarySC}
\end{center}
\end{table}
\subsubsection{Master integrals}
In the following, we present the results for the masters in the hard, collinear 
and soft regions while restricting ourselves to those  
which are explicitly involved in the calculation of the integrated antennae
with two quark flavours.

\subsubsection*{a) The hard region}
\label{hard}

The master integrals defined above were computed in
the hard region mainly with the differential equations
technique~\cite{gr,kotikov,remiddi}. The only masters that were calculated
directly are $I_1,\; I_2$, and $I_{14}$. Since the dependence
of the integrands on $x_1$ and $x_2$ is via the constraints,
$C_1$ and $C_2$, as shown in eq.~(\ref{constarints}),
we derive the differential equations for each master integral by employing
the following operators at the integrand level:
\begin{eqnarray}
\frac{\partial}{\partial x_1} &=& \frac{\partial C_1}{\partial x_1}
\frac{\partial}{\partial C_1} \,+
 \,\frac{\partial C_2}{\partial x_1} \frac{\partial}{\partial C_2} \, \,,\\
\frac{\partial}{\partial x_2} &=& \frac{\partial C_1}{\partial x_2} \frac{\partial}{\partial C_1} \,+
 \,\frac{\partial C_2}{\partial x_2} \frac{\partial}{\partial C_2} \, \,.
\end{eqnarray}
The boundary conditions required for the solution of the differential
equations are either obtained from self-consistency conditions on the
integrals, or by explicit evaluation in the collinear or soft limits.
The solution of the system of differential equations yields two-dimensional
generalized harmonic polylogarithms
(GHPL,~\cite{Gehrmann:2000zt, Gehrmann:2001jv}) of up to weight two,
or products of weight one harmonic polylogarithms 
(HPL,~\cite{Remiddi:1999ew}) of argument $x_1$ or $x_2$.
The definition of the HPL and GHPL functions involved in the solution of the master
integrals in the hard region is recalled below:
\begin{eqnarray}
\H(1,x)	&=& - \ln (1-x)	\nonumber	\\
\H(0,x)	&=& \ln \, x	\nonumber	\\
\H(-1,x)	&=& \ln (1+x)	\nonumber	
\end{eqnarray}
The harmonic polylogarithms of higher weight are defined recursively 
(we group weights into vectors, $b_1, \dotsc, b_w = \vec{b}$):
\begin{equation}
\H(\vec{0}_w, x)	= \frac{1}{w!} \ln^w x
\end{equation}
while, if $\vec{a} = (a,\vec{b}) \neq \vec{0_w}$
\begin{equation}
\H(a, \vec{b}, x )	= \int_0^x \d z f(a, z) H( \vec{b}, z )		\nonumber	
\end{equation}
with weight functions
\begin{equation}
\begin{split}
	f(1,z)		&=	\frac{1}{1-z},	\\
	f(0,z)		&=	\frac{1}{z},		\\
	f(-1,z)	&=	\frac{1}{1+z}.
\end{split}
\end{equation}
This results in the derivative formula
\begin{equation*}
	\frac{ \dd }{ \dd x } H(a_1, \vec{b}, x) = f(a_1,x) H(\vec{b}, x)
\end{equation*}
The two-dimensional generalized harmonic polylogarithms are defined in a 
very similar fashion:
\begin{eqnarray}
 \G(0,y) &=& \ln\, y  \ ,                               \nonumber\\
 \G(1,y) &=& \ln\, (1-y) \ ,                            \nonumber\\
 \G(-1,y) &=& \ln\, (1+y) \ ,                            \nonumber\\
 \G(-z,y) &=& \ln\left( 1 + \frac{y}{z} \right) \ .
\label{eq:w1list}
\end{eqnarray}
For weight $w>1$ we have 
\begin{equation}
  \G(\vec{0}_w,y) = \frac{1}{w!} \ln^w{y} \ ,
\label{eq:defh0}
\end{equation}
\begin{equation}
  \G(\vec{a},y) = \int_0^y \d y' \ \g(a,y') \ \G(\vec{b},y') \ .
\label{eq:defn0}
\end{equation}
where
\begin{equation}
\g(a,y) = \frac{1}{y-a}\,.
\end{equation}
and we have for the derivatives
\begin{equation}
\frac{\dd }{\dd y} \G(\vec{a},y) = \g(a,y) \G(\vec{b},y) \, .
\label{eq:derive}
\end{equation}
As can be seen above, two-dimensional generalized harmonic polylogarithms 
reduce to harmonic polylogarithms if their weights are only $0$, $1$, $-1$, 
with
\begin{equation}
\G( \vec{a}, x) = 
	\begin{cases}
		- \H( \vec{a}, x) 	& \text{if $\vec{a}$ contains an odd number of $1$} \\
		\H( \vec{a}, x)	& \text{else}.
	\end{cases}
\end{equation}
In the results for the hard region below, the GHPL's will have $x_1$ as their 
argument and $0$, $\pm 1$ and $-x_2$ as their weights, the conversion to HPL's 
has been inserted where possible.\\
An important check on our results in this region is to compare the
soft and collinear limits of the hard result against the ones derived
in the soft and collinear regions calculated by a direct integration
of the phase space. We found agreement on all the powers of
the $\ep$-expansion in the hard region using these two different methods.
The results of the masters expanded to the needed order in $\ep$ are given
below.
The common prefactor
$N_{\Gamma}$ is defined as:
\begin{equation}
N_{\Gamma}= \frac{(4\,\pi)^{-4+2\,\ep}}{\Gamma(2 - 2\,\ep)}\,.
\end{equation}
%%%%%%%%%%%%%%%%%%%%%%%%%%%%%%%%%%%%%%%%%%%%%%%%%%%%%%%%%%%%%%%%%%%%%%%%%%%%
\begin{align}
% MR 10.1.2011: k_a -> k_j, k_b -> k_k
\I{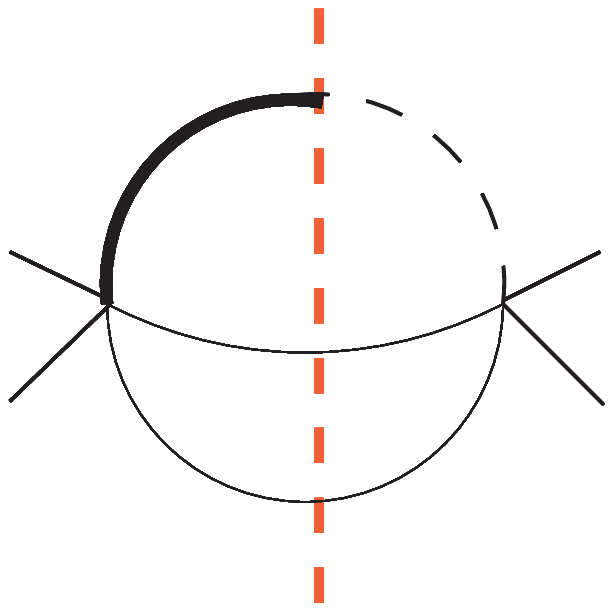} \times k_k \cdot p_1 \;\;&=I_1 = 
(s_{12})^{1 - 2 \ep}
\frac{N_{\Gamma}\, (1-x_1)^{-2 \ep}\,
(1-x_2)^{-2 \ep}}{8\, (x_1+x_2)^2} \,\times
\nonumber \\
&
\Bigl[
-(x_1-1) (x_2-1)^2 (x_2 x_1+2 x_1+x_2)
\nonumber \\
&
+\ep\;\Bigl\{(x_1-1) (x_2 x_1+2 x_1+x_2) \text{H}(0,x_1) (x_2-1)^2
-\frac{1}{2} (x_1-1) \,\times
\nonumber \\
&
\left(5 x_1 x_2^2+3 x_2^2+3 x_1 x_2-7 x_2-12 x_1\right) (x_2-1)
\nonumber \\
&
-2 \left(x_1^2 x_2^3-x_2^3-2 x_1 x_2^2-3 x_1^2 x_2-x_2-2
x_1\right) \text{G}(-x_2,x_1)
\nonumber \\
&
+(x_1+1) (x_2+1)^2 (x_2 x_1-2 x_1-x_2) (\text{H}(-1,x_1)+\text{H}(-1,x_2)
\nonumber \\
&
-\text{H}(0,x_2))+4 (x_1+x_2)^2 \log (2)\Bigr\}
\nonumber \\
&
+\ep^2\;\Bigl\{+(x_1-1) (x_2 x_1+2 x_1+x_2)
(\text{H}(0,-1,x_2)-\text{H}(0,0,x_1)-\text{H}(0,0,x_2)) \,\times
\nonumber \\
& 
(x_2-1)^2-\frac{9}{4} (x_1-1) \left(3 x_1 x_2^2+x_2^2+x_1 x_2-5 x_2-8
x_1\right) (x_2-1)
\nonumber \\
& 
+\frac{1}{2} (x_1-1) \left(5 x_1 x_2^2+3 x_2^2+3 x_1 x_2-7 x_2-12
x_1\right) \text{H}(0,x_1) (x_2-1)
\nonumber \\
& 
+\left(-5 x_1^2 x_2^3+3 x_2^3+8 x_1 x_2^2+15 x_1^2 x_2+7 x_2+12
x_1\right) \text{G}(-x_2,x_1)
\nonumber \\
&
+\left(x_1^2 x_2^3-x_2^3-2 x_1 x_2^2-3 x_1^2 x_2-x_2-2 x_1\right)
(2 \text{G}(0,-x_2,x_1)
\nonumber \\
& 
+2 \text{G}(-x_2,-1,x_1)
+2 \text{G}(-x_2,0,x_1)-4 \text{G}(-x_2,-x_2,x_1)
\nonumber \\
& 
+\text{G}(-x_2,x_1) (2 \text{H}(-1,x_2)-2 \text{H}(0,x_2)))+(x_1+1) (x_2+1) 
\,\times
\nonumber \\
& 
\left(5 x_1 x_2^2-3 x_2^2-3 x_1 x_2-7 x_2-12
x_1\right) \left(\frac{1}{2} \text{H}(-1,x_1)+\frac{1}{2} \text{H}(-1,x_2)
\right.\nonumber \\
& \left.
-\frac{1}{2} \text{H}(0,x_2)\right)+\left(-x_1^2 x_2^3+x_2^3+2 x_1 x_2^2+6
x_2^2+3 x_1^2 x_2+12 x_1 x_2
\right.\nonumber \\
& \left.
+x_2+6 x_1^2+2 x_1\right)
(\text{G}(1,-x_2,x_1)+\text{H}(-1,x_2) \text{H}(1,x_1)+\text{H}(1,-1,x_1)
\nonumber \\
& 
+\text{H}(1,-1,x_2))+\left(x_1^2 x_2^3-x_2^3-2 x_1 x_2^2-6 x_2^2-3 x_1^2
x_2-12 x_1 x_2
\right.\nonumber \\
& \left.
-x_2-6 x_1^2-2 x_1\right) (\text{H}(0,x_2) \text{H}(1,x_1)+\log
(2) \text{H}(1,x_1)+\text{H}(1,0,x_2)
\nonumber \\
& 
+\text{H}(1,x_2) \log (2))+(x_1+x_2)^2 \left(-4 \log
(2) \text{H}(0,x_1)-4 \log ^2(2)
\right.\nonumber \\
& \left.
-4 \text{H}(0,x_2) \log (2)\right)+(x_1+1) (x_2+1)^2 (x_2 x_1-2 x_1-x_2) \,\times
\nonumber \\
&
\left(3 \text{G}(-1,-x_2,x_1)+\text{H}(0,x_1) \text{H}(0,x_2)
-2 \text{H}(-1,-1,x_1)
\right.\nonumber \\
& \left.
-2 \text{H}(-1,-1,x_2)-\text{H}(-1,0,x_1)+2 \text{H}(-1,0,x_2)-\text{H}(0,-1,x_1)
\right.\nonumber \\
& \left.
+\text{H}(-1,x_2) (\log (2)-\text{H}(0,x_1))+\text{H}(-1,x_1)
(-2 \text{H}(-1,x_2)
\right.\nonumber \\
& \left.
+2 \text{H}(0,x_2)+\log (2))+\frac{\pi ^2}{3}\right)
\nonumber \\
&
-2 (x_1+x_2) \left(x_1x_2^2-5 x_2-6 x_1\right) \log (2)\Bigr\}
+ \mathcal{O}(\ep^3)\; \Bigr]\,.
\end{align}

\begin{align}
\I{1} \;\;&=I_2 =
(s_{12})^{- 2 \ep}
 \frac{N_{\Gamma}\, (1-x_1)\,^{-2 \ep}
(1-x_2)\,^{-2 \ep} }{(1-3\, \ep)\, \,(x_1+x_2)\,}
\nonumber \\
&
\Bigl[
+(x_1-1)\, (x_2-1)\,
+\,\ep\;\Bigl\{2 \log (2)\, (x_1+x_2)\,+2
(x_1 x_2+1)\, \text{G}(-x_2,x_1)\,
\nonumber \\
&
-(x_1-1)\, (x_2-1)\, \text{\,H}(0,x_1)\,+(x_1+1)\,
(x_2+1)\, (-\text{\,H}(-1,x_1)\,
\nonumber \\
&
-\text{\,H}(-1,x_2)\,+\text{\,H }(0,x_2)\,)\,\Bigr\}
\nonumber \\
&
+\,\ep^2\;\Bigl\{(x_1 x_2+1)\,
(-2 \text{G}(0,-x_2,x_1)\,-2 \text{G}(-x_2,-1,x_1)\,-2 \text{G}(-x_2,0,x_1)\,
\nonumber \\
&
+4\, \text{G}(-x_2,-x_2,x_1)\,+\text{G}(-x_2,x_1)\,
(2 \text{\,H}(0,x_2)\,-2 \text{\,H}(-1,x_2)\,)\,)\,
\nonumber \\
&
+(x_1-1)\, (x_2-1)\,
(-\text{\,H}(0,-1,x_2)\,+\text{\,H}(0,0,x_1)\,+\text{\,H}(0,0,x_2)\,)\,
\nonumber \\
&
+(x_2 x_1+3\,x_1+3\, x_2+1)\,
(\text{G}(1,-x_2,x_1)\,+\text{\,H}(-1,x_2)\, \text{\,H}(1,x_1)\,
\nonumber \\
&
+\text{\,H}(1,-1,x_1)\,+\text{\,H}(1,-1,x_2)\,)\,+(x_1+1)\,
(x_2+1)\, \,\times 
\nonumber \\
&
\left(-3\, \text{G}(-1,-x_2,x_1)\,-\text{\,H}(0,x_1)\, \text{\,H}(0,x_2)\,+2 \text{\,H}(-1,-1,x_1)\,+2 \text{\,H}(-1,-1,x_2)\,
\right. \nonumber \\
& \left.
+\text{\,H}(-1,0,x_1)\,-2 \text{\,H}(-1,0,x_2)\,+\text{\,H}(0,-1,x_1)\,
+\text{\,H}(-1,x_2)\,(2 \text{\,H}(-1,x_1)\,
\right.\nonumber \\
& \left.
+\text{\,H}(0,x_1)\,-\log (2)\,)\,+\text{\,H}(-1,x_1)\,
(-2 \text{\,H}(0,x_2)\,-\log (2)\,)\,-\frac{\pi ^2}{3\,}\right)\,
\nonumber \\
&
+(-x_2 x_1-3\, x_1-3\,x_2-1)\, (\text{\,H}(0,x_2)\, \text{\,H}(1,x_1)\,+\log
(2)\, \text{\,H}(1,x_1)\,
\nonumber \\
&
+\text{\,H}(1,0,x_2)\,
%\nonumber \\
%&
+\text{\,H}(1,x_2)\, \log(2)\,)\,+(x_1+x_2)\, \left(-2 \log
(2)\, \text{\,H}(0,x_1)\,
\right.\nonumber \\
& \left.
-2 \log^2(2)\,
-2 \text{\,H}(0,x_2)\, \log (2)\,\right)\,\Bigr\} + \mathcal{O}(\ep^3)\;\Bigr]
\,.
\end{align}

\begin{align}
% MR 10.1.2011: k_a -> k_j, k_b -> k_k
\I{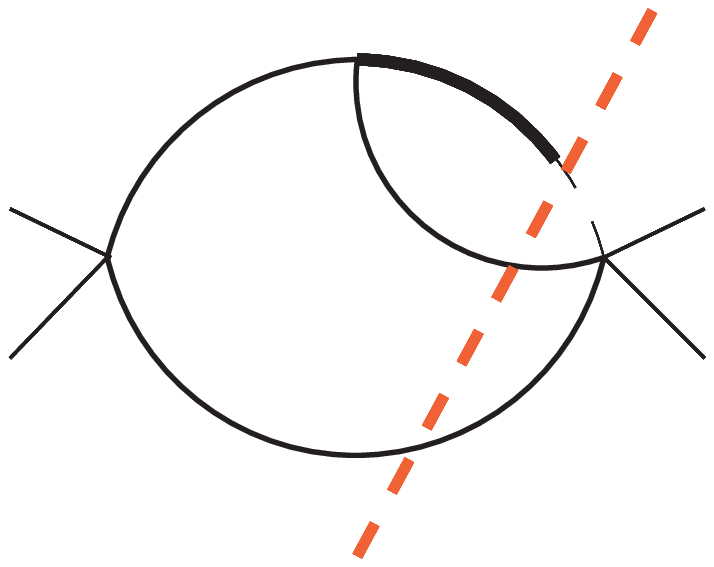} &\times (-(k_j \cdot p_1)(k_j \cdot p_3))  \;\; =I_3
= 
(s_{12})^{1 - 2 \ep}
\frac{(1-2 \ep) \,N_{\Gamma}\, (1-x_1)^{-2 \ep}
(1-x_2)^{-2 \ep} }{8 (x_1+x_2)^2} \,\times
\nonumber \\
&
\Bigl[(x_1-1) (5 x_1-1) \text{H}(0,x_2) x_2^2-2 x_1 (x_1-x_2) \text{H}(0,x_1)
 \text{H}(0,x_2) x_2^2
\nonumber \\
&
-(x_1-1) (5 x_1-x_2) (x_2-1) x_2+x_1 (x_2-1) 
\left(-4
 x_2^2+x_1 x_2+2 x_2+x_1\right) \text{H}(0,x_1)
\nonumber \\
&
+\ep \,\Bigl\{-\left(-x_1^2+4 x_2
 x_1-6 x_1+1\right) \text{H}(0,x_1) \text{H}(0,x_2) x_2^2
+(x_1+1) (5 x_1+1)\,\times
\nonumber \\
&
 (-\text{H}(-1,x_1) \text{H}(0,x_2)-\text{H}(-1,0,x_1)
-\text{H}(0,-1,x_2)) x_2^2
+(x_1-1) (5 x_1-1)\,\times
\nonumber \\
&
 (2 \text{H}(0,1,x_2)-2 \text{H}(0,x_2) \text{H}(1,x_2)) x_2^2+x_1
 (12 \text{H}(0,x_2) \log (2)-12 \text{H}(0,x_1) \log (2)) x_2^2
\nonumber \\
&
+x_1
 (x_1-x_2) \left(-2 \text{G}(0,-x_2,0,x_1)-2 \text{G}(0,-x_2,x_1) 
\text{H}(0,x_2)
\right. \nonumber \\
& \left.
-4 \text{G}(-x_2,0,x_1) \text{H}(0,x_2)+\text{H}(0,x_2) \left(2 \text{H}(0,-1,x_1)+6 \text{H}(0,0,x_1)-4 \text{H}(0,1,x_1)
\right. \right.\nonumber \\
& \left. \left.
+\frac{5 \pi
 ^2}{6}\right)+\text{H}(0,x_1) \left(\text{H}(0,x_2)
 (4 \text{H}(1,x_1)+4 \text{H}(1,x_2))+2 \text{H}(0,-1,x_2)
\right.\right. \nonumber \\
& \left.\left.
-4 \text{H}(0,1,x_2)+\frac{\pi
 ^2}{2}\right)+2 \text{H}(0,-1,0,x_1)+2 \text{H}(0,-1,0,x_2)-2 \text{H}(0,0,0,x_2)
\right.\nonumber \\
& \left.
+3 \zeta(3)\right) x_2^2-(x_1-1) (23 x_1-5 x_2) (x_2-1) x_2-2 (5 x_1-x_2) (x_1
 x_2+1) \,\times
\nonumber \\
&
\text{G}(-x_2,x_1) x_2+(x_1+1) (5 x_1-x_2) (x_2+1)
 (\text{H}(-1,x_1)+\text{H}(-1,x_2)) x_2
\nonumber \\
&
+\left(4 x_2 x_1^2-6 x_1^2+x_2^2
 x_1-15 x_2 x_1-4 x_1+x_2^2+3 x_2\right) \text{H}(0,x_2) x_2
\nonumber \\
&
-2 (5 x_1-x_2)
 (x_1+x_2) \log (2) x_2
+\frac{1}{12} \pi ^2 \left(20 x_1 x_2^3-20 x_1^2 x_2^2+12 x_1 x_2^2-3 x_2^2
\right.\nonumber \\
& \left.
+10 x_1 x_2+5 x_1^2\right)+2 x_1 \left(-4 x_2^3+x_1 x_2^2-2
x_2-x_1\right) \text{G}(0,-x_2,x_1)
\nonumber \\
&
+\left(-4 x_1 x_2^3+6 x_1^2 x_2^2+12 x_1 x_2^2+x_2^2-2 x_1
x_2-x_1^2\right) \text{G}(-x_2,0,x_1)
\nonumber \\
&
-(x_2-1) \left(-6 x_2 x_1^2-2 x_1^2+9 x_2^2 x_1+2 x_2
x_1-x_2^2\right) \text{H}(0,x_1)
\nonumber \\
&
+\left(-4 x_1 x_2^3+6 x_1^2 x_2^2-12 x_1 x_2^2+x_2^2-2 x_1
x_2-x_1^2\right) \text{G}(-x_2,x_1) \text{H}(0,x_2)
\nonumber \\
&
+x_1 (x_2+1) \left(-4 x_2^2+x_1 x_2-2 x_2-x_1\right) \,\times
\nonumber \\
&
(-\text{H}(-1,x_2) \text{H}(0,x_1)-\text{H}(-1,0,x_2)
-\text{H}(0,-1,x_1))+x_1 (x_2-1) \,\times
\nonumber \\
&
\left(-4 x_2^2+x_1 x_2+2 x_2+x_1\right)
(-2 \text{H}(0,x_1) \text{H}(1,x_1)-3 \text{H}(0,0,x_1)
\nonumber \\
&
+\text{H}(0,0,x_2)+2 \text{H}(0,1,x_1))\Bigr\} + \mathcal{O}(\ep^2)\;\Bigr]\,.
\end{align}

\begin{align}
% MR 10.1.2011: k_a -> k_j, k_b -> k_k
\I{6}\times k_j \cdot p_1 \;\;& =I_4 =
(s_{12})^{ - 2 \ep}
 \frac{(1-2 \ep)\, \,N_{\Gamma}\,
(1-x_1)\,^{-2 \ep}(1-x_2)\,^{-2 \ep} }{4 (x_1+x_2)\,^2} \,\times
\nonumber \\
&
\Bigl[6 (x_1-1)\, (x_2-1)\, x_2-2 (x_1-1)\, (x_2+2)\, \text{\,H}(0,x_2)\,
x_2
\nonumber \\
&
+\text{\,H}(0,x_1)\, (2 x_2 \text{\,H}(0,x_2)\,-2 (x_2-1)\, (x_2
x_1-x_1+x_2)\,)\,
\nonumber \\
&
+\ep \;\Bigl\{12 x_2 \log (2)\, (x_1+x_2)\,+\frac{1}{6} \pi ^2 \left(8
x_1 x_2^2-2 x_2^2+4 x_1 x_2-11 x_2+5 x_1\right)\,
\nonumber \\
&
+2\,x_2 \text{\,G}(0,-x_2,0,x_1)\,+\text{\,H}(-1,x_2)\, (2 (x_2+1)\, (x_2
x_1+x_1-x_2)\, \text{\,H}(0,x_1)\,
\nonumber \\
&
-6 (x_1+1)\, x_2(x_2+1)\,)\,+\text{\,G}(0,-x_2,x_1)\, \left(2
x_2 \text{\,H}(0,x_2)\,
\right. \nonumber \\
&\left.
-4 \left(x_1x_2^2-x_2+x_1\right)\,\right)\,+\text{\,G}(-x_2,0,x_1)\, \left(4
x_2 \text{\,H}(0,x_2)\,
\right. \nonumber \\
& \left.
-2 \left(2 x_1 x_2^2+2 x_2^2-4 x_1 x_2-3
x_2+x_1\right)\,\right)\,
\nonumber \\
&
+\text{\,H}(-1,x_1)\, (2 (x_1+1)\, (x_2-2)\,
x_2 \text{\,H}(0,x_2)\,-6 (x_1+1)\, x_2 (x_2+1)\,)\,
\nonumber \\
&
+\text{\,G}(-x_2,x_1)\, \left(12 x_2\,(x_1 x_2+1)\,
\right. \nonumber \\
& \left.
-2 \left(2 x_1 x_2^2-2 x_2^2+4 x_1 x_2-3
x_2+x_1\right)\, \text{\,H}(0,x_2)\,\right)\,
\nonumber \\
&
+2 (x_1+1)\, (x_2-2)\,
x_2 \text{\,H}(-1,0,x_1)\,+2 (x_2+1)\, (x_2 x_1+x_1-x_2)\, \times
\nonumber \\
&
\text{\,H}(-1,0,x_2)\,+2\,(x_2+1)\, (x_2 x_1+x_1-x_2)\,
\text{\,H}(0,-1,x_1)\,
\nonumber \\
&
+2 (x_1+1)\, (x_2-2)\,
x_2 \text{\,H}(0,-1,x_2)\,+6 (x_2-1)\, (x_2 x_1-x_1+x_2)\, \times
\nonumber \\
&
\text{\,H}(0,0,x_1)\,-2
(x_2-1)\, (x_2 x_1-x_1+x_2)\, \text{\,H}(0,0,x_2)\,
\nonumber \\
&
-4 (x_2-1)\, (x_2
x_1-x_1+x_2)\, \text{\,H}(0,1,x_1)\,-4 (x_1-1)\, x_2 (x_2+2)\, \times
\nonumber \\
&
\text{\,H}(0,1,x_2)\,-2
x_2 \text{\,H}(0,-1,0,x_1)\,-2 x_2 \text{\,H}(0,-1,0,x_2)\,
\nonumber \\
&
+2\,x_2 \text{\,H}(0,0,0,x_2)\,+\text{\,H}(0,x_1)\, \left(-2 \text{\,H}(0,-1,x_2)\,
x_2+4 \text{\,H}(0,1,x_2)\, x_2
\right. \nonumber \\
& \left.
-4 (2 x_1-x_2)\, \log (2)\, x_2-\frac{\pi ^2
x_2}{2}-4 (x_2-1)\, (2 x_2 x_1-x_1-x_2)\,
\right.\nonumber \\
& \left.
+4 (x_2-1)\,
(x_2\,x_1-x_1+x_2)\, \text{\,H}(1,x_1)\,+\text{\,H}(0,x_2)\, \times
\right. \nonumber \\
& \left.
(-2 x_2 (x_2 x_1-2
x_1+x_2+2)\,-4 x_2 \text{\,H}(1,x_1)\,-4
x_2 \text{\,H}(1,x_2)\,)\,\right)\,
\nonumber \\
&
+\text{\,H}(0,x_2)\, \left(4 (x_1 x_2+2 x_2+3)\,
x_2+4 (x_1-1)\, (x_2+2)\, \text{\,H}(1,x_2)\, x_2
\right. \nonumber \\
& \left.
-2 \text{\,H}(0,-1,x_1)\,
x_2-6 \text{\,H}(0,0,x_1)\, x_2
\right. \nonumber \\
& \left.
+4 \text{\,H }(0,1,x_1)\, x_2+4 (2 x_1-x_2)\, \log (2)\, x_2-\frac{5 \pi ^2
x_2}{6}\right)\,
\nonumber \\
&
+x_2 (26 x_2 x_1-26 x_1-26 x_2-3 \zeta (3)\,+26)\,\Bigr\} + \mathcal{O}(\ep^2)\,\Bigr]\,.
\end{align}

\begin{align}
% MR 10.1.2011: k_a -> k_j, k_b -> k_k
\I{6}\times (- k_j \cdot p_3) \;\;& =I_5 = 
(s_{12})^{ - 2 \ep}
\frac{(1-2 \,\ep)\,\, N_{\Gamma} \,(1-x_1)^{-2 \,\ep}
(1-x_2)^{-2 \,\ep}}{2 \,(x_1+x_2)} \,\times
\nonumber \\
&
\Bigl[x_1 (x_2-1)\,\, \text{\,H}(0,x_1)\,\,-(x_1-1)\,\,
x_2 \text{\,H}(0,x_2)\,
\nonumber \\
&
+\,\ep \, \Bigl\{\frac{1}{12} \pi ^2 (2 x_2 x_1+5 x_1-3 x_2)\,\,
-2x_1 \text{\,G}(0,-x_2,x_1)\,\,
\nonumber \\
&
+(2 x_2 x_1-x_1+x_2)\,\, \text{\,G}(-x_2,0,x_1)\,\,
+(-2 x_2x_1-x_1+x_2)\,\times
\nonumber \\
&
\text{\,G}(-x_2,x_1)\,\, \text{\,H}(0,x_2)\,\,-(x_1+1)\,\,
x_2 \text{\,H}(-1,0,x_1)\,\,
\nonumber \\
&
+x_1 (x_2+1)\,\, \text{\,H}(-1,0,x_2)\,\,
+ x_1 (x_2+1)\,\,\text{\,H}(0,-1,x_1)\,\,
\nonumber \\
&
-(x_1+1)\,\, x_2 \text{\,H}(0,-1,x_2)\,\,
-3 x_1(x_2-1)\,\, \text{\,H}(0,0,x_1)\,\,
\nonumber \\
&
+x_1(x_2-1)\,\, \text{\,H}(0,0,x_2)\,
+2 x_1(x_2-1)\,\, \text{\,H}(0,1,x_1)\,\,
\nonumber \\
&
-2 (x_1-1)\,\,x_2 \text{\,H}(0,1,x_2)\,\,
%\nonumber \\
%&
+\text{\,H}(0,x_1)\,\, (2 x_1\,(x_2-1)\,
\nonumber \\
&
-2 x_1 \text{\,H}(1,x_1)\,\, (x_2-1)\,\,
+x_1 (x_2+1)\,\, \text{\,H}(-1,x_2)\,\,
\nonumber \\
&
+(x_1-1)\,\,x_2 \text{\,H}(0,x_2)\,\,-2 x_1 x_2 \log (2)\,\,)\,\,
\nonumber \\
&
+\text{\,H}(0,x_2)\,\, (-2 (x_1-1)\,\,
x_2-(x_1+1)\,\, \text{\,H}(-1,x_1)\,\, x_2
\nonumber \\
&
+2 (x_1-1)\,\, \text{\,H}(1,x_2)\,\, x_2
+2 x_1 x_2\,\log (2))\, \Bigr\} + \mathcal{O}(\ep^2)\;\Bigr]\,.
\end{align}

\begin{align}
\I{6} \;\;& =I_6 = 
(s_{12})^{-1 - 2 \ep}
\frac{(1-2 \,\ep) \,N_{\Gamma}\, (1-x_1)^{-2 \,\ep}
(1-x_2)^{-2 \,\ep}}{x_1+x_2} \,\times
\nonumber \\
&
\Bigl[\text{\,H}(0,x_1) \text{\,H}(0,x_2)
+\,\ep \,\Bigl\{\text{\,G}(0,-x_2,0,x_1)
+\text{\,G}(0,-x_2,x_1) \text{\,H}(0,x_2)
\nonumber \\
&
+2 \text{\,G}(-x_2,0,x_1) \text{\,H}(0,x_2)
+\text{\,H}(0,x_2) \left(-\text{\,H}(0,-1,x_1)
-3 \text{\,H}(0,0,x_1)
\right. \nonumber \\
& \left. 
+2 \text{\,H}(0,1,x_1)-\frac{5 \pi^2}{12}\right)
+\text{\,H}(0,x_1) \left(\text{\,H}(0,x_2)
(-2 \text{\,H}(1,x_1)-2 \text{\,H}(1,x_2))
\right. \nonumber \\
& \left.
-\text{\,H}(0,-1,x_2)
+2 \text{\,H}(0,1,x_2)-\frac{\pi^2}{4}\right)-\text{\,H}(0,-1,0,x_1)
-\text{\,H}(0,-1,0,x_2)
\nonumber \\
&
+\text{\,H }(0,0,0,x_2)
-\frac{3 \zeta (3)}{2}\Bigr\} + \mathcal{O}(\ep^2)\, \Bigr]\,.
\end{align}

\begin{align}
\I{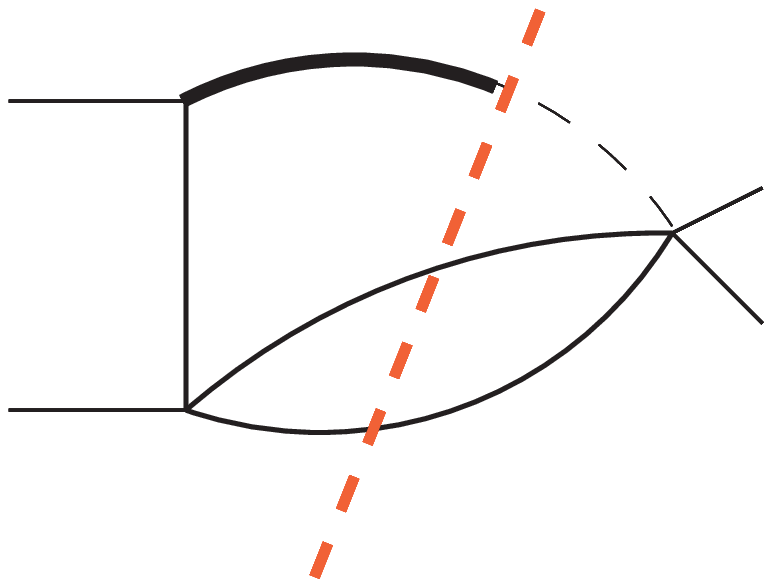} \;\;& =I_7 = 
(s_{12})^{-1 - 2 \ep}
\frac{(1-2 \,\ep)\, N_{\Gamma} (1-x_1)\,^{-2 \,\ep}
(1-x_2)\,^{-2 \,\ep}}{x_2} \,\times 
\nonumber \\
&
\Bigl[\text{\,G}(-x_2,x_1)\,-\text{\,H}(-1,x_1)\,+\text{\,H}(0,x_2)\, 
+\ep\,\Bigl\{-3 \text{\,G}(-1,-x_2,x_1)\,
\nonumber \\
&
-\text{\,G}(0,-x_2,x_1)\,
-\text{\,G}(1,-x_2,x_1)\,-\text{\,G}(-x_2,-1,x_1)\,
-\text{\,G}(-x_2,0,x_1)\,
\nonumber \\
&
+2 \text{\,G}(-x_2,-x_2,x_1)\,
+\text{\,G}(-x_2,x_1)\,
(-\text{\,H}(-1,x_2)\,+\text{\,H}(0,x_2)\,+2)\,
\nonumber \\
&
-\text{\,H}(-1,x_2)\, \text{\,H}(1,x_1)\,
+\text{\,H}(0,x_2)\,
(-\text{\,H}(0,x_1)\,+\text{\,H}(1,x_1)\,-2 \text{\,H}(1,x_2)\,+2)\,
\nonumber \\
&
+2 \text{\,H}(-1,-1,x_1)\,
+\text{\,H}(-1,0,x_1)\,-2 \text{\,H}(-1,1,x_1)\,
\nonumber \\
&
+\text{\,H}(0,-1,x_1)\,-\text{\,H}(0,-1,x_2)\,+2 \text{\,H}(0,1,x_2)\,
-3 \text{\,H}(1,-1,x_1)\,
\nonumber \\
&
+\text{\,H}(-1,x_1)\,
(-2 \text{\,H}(0,x_2)\,+2 \text{\,H}(1,x_1)\,+\log (2)\,-2)\,
\nonumber \\
&
+\text{\,H }(1,x_1)\, \log(2)\,-\frac{\pi ^2}{4}
\Bigr\} + \mathcal{O}(\ep^2) \,\Bigr]\,.
\end{align}

\begin{align}
\I{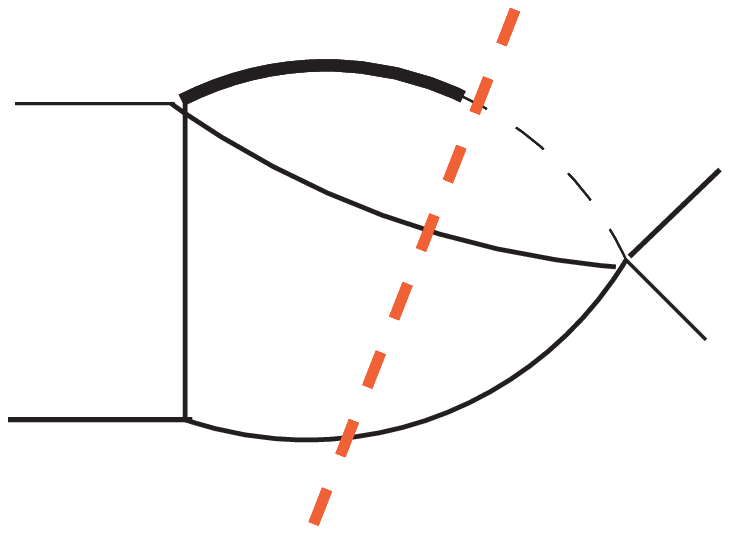} \;\;& =I_8 = 
(s_{12})^{-1 - 2 \ep}
\frac{(1-2 \,\ep)\, N_{\Gamma}\, (1-x_1)\,^{-2 \,\ep}
(1-x_2)\,^{-2 \,\ep}}{x_1} \,\times
\nonumber \\
&
\Bigl[
+\frac{1}{\ep}\Bigl\{\text{G}(-x_2,x_1)\,-\text{\,H}(-1,x_1)\,\Bigr\}
\nonumber\\
&
-3 \text{G}(-1,-x_2,x_1)\,+2 \text{G}(0,-x_2,x_1)\,-\text{G}(1,-x_2,x_1)\,
-\text{G}(-x_2,-1,x_1)\,
\nonumber \\
&
-\text{G}(-x_2,0,x_1)\,+2 \text{G}(-x_2,-x_2,x_1)\,
\nonumber \\
&
+\text{G}(-x_2,x_1)\,
(\text{\,H}(0,x_2)\,-\text{\,H}(-1,x_2)\,)\,-\text{\,H}(-1,x_2)\,
\text{\,H}(1,x_1)\,
\nonumber \\
&
+\text{\,H}(0,x_2)\, \text{\,H}(1,x_1)\,
+2 \text{\,H}(-1,-1,x_1)\,+\text{\,H}(-1,0,x_1)\,-2 \text{\,H}(-1,1,x_1)\,
\nonumber \\
&
-2 \text{\,H}(0,-1,x_1)\,
-3 \text{\,H}(1,-1,x_1)\,+\text{\,H}(-1,x_1)\, (2 \text{\,H}(1,x_1)\,+\log
(2)\,)\,
\nonumber \\
&
+\text{\,H}(1,x_1)\, \log (2)\,+ \mathcal{O}(\ep)\,\Bigr]\,.
\end{align}

\begin{align}
\I{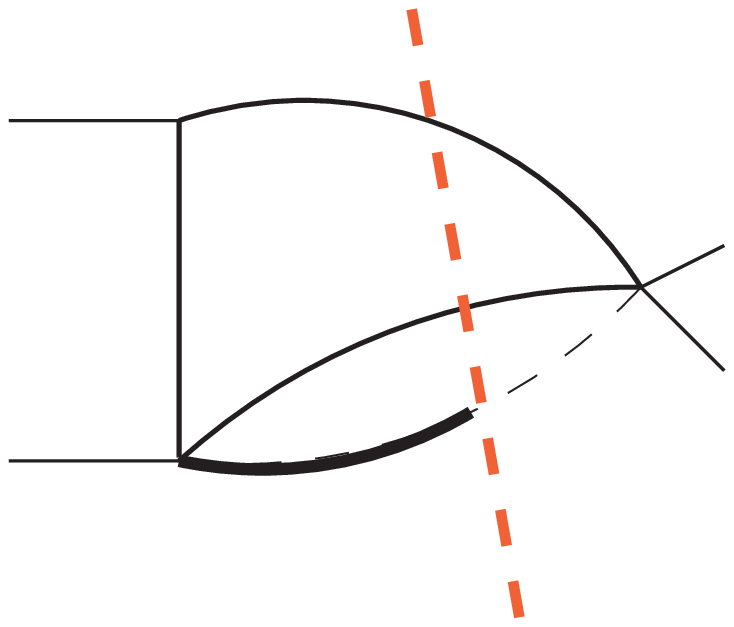} \;\;& =I_9 = 
(s_{12})^{-1 - 2 \ep}
\frac{(1-2 \,\ep)\, N_{\Gamma}\, (1-x_1)\,^{-2 \,\ep} 
(1-x_2)\,^{-2 \,\ep}}{x_2} \,\times
\nonumber \\
&
\Bigl[
+ \frac{1}{\,\ep}\Bigl\{\text{\,G}(-x_2,x_1)\,-\text{\,H}(-1,x_2)\,-\text{\,H}(0,x_1)\,+\text{\,H}(0,x_2)\,\Bigr\}
\nonumber \\
&
-4 \text{\,G}(0,-x_2,x_1)\,+2 \text{\,G}(1,-x_2,x_1)\,
-\text{\,G}(-x_2,-1,x_1)\,-\text{\,G}(-x_2,0,x_1)\,
\nonumber \\
&
+2 \text{\,G}(-x_2,-x_2,x_1)\,+\text{\,G}(-x_2,x_1)\,
(\text{\,H}(0,x_2)\,-\text{\,H}(-1,x_2)\,)\,
\nonumber \\
&
+\text{\,H}(0,x_1)\,(2 \text{\,H}(1,x_1)\,
-3 \text{\,H}(0,x_2)\,)\,
+\text{\,H}(0,x_2)\,
(-2 \text{\,H}(1,x_1)\,-2 \text{\,H}(1,x_2)\,)\,
\nonumber \\
&
+2 \text{\,H}(-1,-1,x_2)\,
-2 \text{\,H}(-1,0,x_2)\,-2 \text{\,H}(-1,1,x_2)\,+\text{\,H}(0,-1,x_1)\,
\nonumber \\
&
-3 \text{\,H}(0,-1,x_2)\,+3 \text{\,H}(0,0,x_1)\,
+3 \text{\,H}(0,0,x_2)\,-2 \text{\,H}(0,1,x_1)\,+2 \text{\,H}(0,1,x_2)\,
\nonumber \\
&
-3 \text{\,H}(1,-1,x_2)\,+3 \text{\,H}(1,0,x_2)\,
+\text{\,H}(-1,x_2)\,
(3 \text{\,H}(0,x_1)\,+2 \text{\,H}(1,x_1)\,
\nonumber \\
&
+2 \text{\,H}(1,x_2)\,+\log(2)\,)\,+\text{\,H}(1,x_2)\, \log (2)\,
+\frac{\pi ^2}{2} + \mathcal{O}(\,\ep)\,\Bigr]\,.
\end{align}

\begin{align}
\I{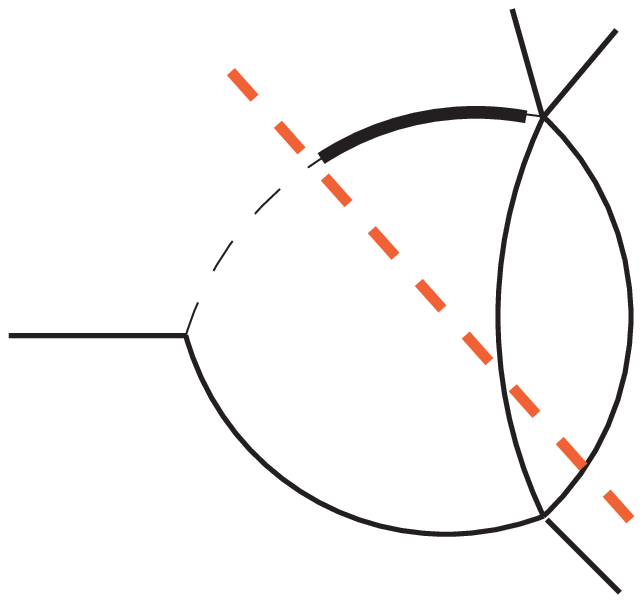} \;\;&=I_{10} =  
(s_{12})^{-1 - 2 \ep}
\frac{(1-2 \,\ep)\, N_{\Gamma}\, (1-x_1)\,^{-2 \,\ep}
(1-x_2)\,^{-2 \,\ep}}{x_1} \, \times
\nonumber \\ 
&\Bigl[ \,\text{G}(-x_2,x_1)\,-\,\text{H}(-1,x_2)\,+\,\text{H}(0,x_2)\,
+\,\ep \;\Bigl\{-\,\text{G}(0,-x_2,x_1)\,
\nonumber \\
&
+2 \,\text{G}(1,-x_2,x_1)\,
%\right. \right. \nonumber \\&
-\,\text{G}(-x_2,-1,x_1)\,-\,\text{G}(-x_2,0,x_1)\,+2 \,\text{G}(-x_2,-x_2,x_1)\,
\nonumber \\
&
-\,\text{H}(0,x_1)\, \,\text{H}(0,x_2)\,
+\,\text{G}(-x_2,x_1)\, (-\,\text{H}(-1,x_2)\,+\,\text{H}(0,x_2)\,+2)\,
\nonumber \\
&
+\,\text{H}(0,x_2)\, (-2 \,\text{H}(1,x_1)\,
-2 \,\text{H}(1,x_2)\,+2)\,
+2 \,\text{H}(-1,-1,x_2)\,
\nonumber \\
&
-2 \,\text{H}(-1,0,x_2)\,-2 \,\text{H}(-1,1,x_2)\,
+2 \,\text{H}(0,1,x_2)\,-3 \,\text{H}(1,-1,x_2)\,
%%%%%\nonumber \\ & \left. \left.
\nonumber \\
&
+3 \,\text{H}(1,0,x_2)\,
+\,\text{H}(-1,x_2)\, (\,\text{H}(0,x_1)\,
+2 \,\text{H}(1,x_1)\,
+2 \,\text{H}(1,x_2)\,
\nonumber \\
&
+\log (2)\,-2)\,
+\,\text{H}(1,x_2)\, \log (2)\,-\frac{\pi ^2}{4}\Bigr\} + \mathcal{O}(\,\ep^2)\,\Bigr]\,. 
\end{align}

\begin{align}
\I{6} \;\;& =I_{15} = 
(s_{12})^{-2 - 2 \ep}
\frac{2^{-4 \ep} \,(1-2 \ep) \,N_{\Gamma}\,
\,(1-x_1)^{-2 \ep} \,(1-x_2)^{-2 \ep}}{x_1 \,x_2} \,\times
\nonumber \\
&
\Bigl[
+\frac{1}{2\, \ep^2}
+\frac{1}{\ep}\;\Bigl\{\text{\,H}\,(-1,x_1)+\text{\,H}\,(-1,x_2)-\text{\,H}\,(0,x_1)-\text{\,H}\,(0,x_2)\Bigr\}
\nonumber \\
&
-2 \text{\,H}\,(1,x_1)^2+4 \log\,(2) \text{\,H}\,(1,x_1)-2 \text{\,H}\,(1,x_2)^2+3 \text{G}\,(-1,-x_2,x_1)
\nonumber \\
&
-3 \text{G}\,(1,-x_2,x_1)
+\text{\,H}\,(0,x_1)
\,(2 \text{\,H}\,(0,x_2)+2 \text{\,H}\,(1,x_1))
\nonumber \\
&
+\text{\,H}\,(0,x_2)
\,(3 \text{\,H}\,(1,x_1)+2 \text{\,H}\,(1,x_2))-2 \text{\,H}\,(-1,-1,x_1)-2 \text{\,H}\,(-1,-1,x_2)
\nonumber \\
&
-\text{\,H}\,(-1,0,x_1)+2 \text{\,H}\,(-1,0,x_2)+2 \text{\,H}\,(-1,1,x_1)+2 \text{\,H}\,(-1,1,x_2)
\nonumber \\
&
-2 \text{\,H}\,(0,-1,x_1)
-2 \text{\,H}\,(0,-1,x_2)+2 \text{\,H}\,(0,0,x_1)+2 \text{\,H}\,(0,0,x_2)
\nonumber \\
&
-2 \text{\,H}\,(0,1,x_1)-2 \text{\,H}\,(0,1,x_2)-2 \text{\,H}\,(1,-1,x_1)-2 \text{\,H}\,(1,-1,x_2)
\nonumber \\
&
-\text{\,H}\,(1,0,x_1)+2 \text{\,H}\,(1,0,x_2)+4 \text{\,H}\,(1,1,x_1)
+4 \text{\,H}\,(1,1,x_2)
\nonumber \\
&
+\text{\,H}\,(-1,x_1)
\,(-\text{\,H}\,(-1,x_2)+\text{\,H}\,(0,x_2)-2 \text{\,H}\,(1,x_1)+4 \log
\,(2))
\nonumber \\
&
+\text{\,H}\,(-1,x_2)
\,(-2 \text{\,H}\,(0,x_1)-3 \text{\,H}\,(1,x_1)-2 \text{\,H}\,(1,x_2)+4 \log
\,(2))
\nonumber \\
&
+4 \text{\,H}\,(1,x_2) \log \,(2)
+\frac{3 \pi ^2}{4}+ \mathcal{O}(\ep)\,\Bigr]
\end{align}

\begin{align} 
\label{M14hard}
\I{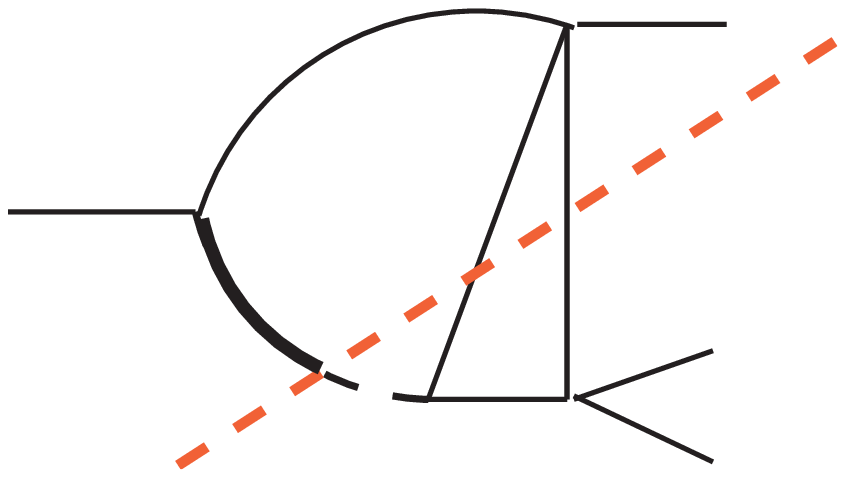} \;\;&=I_{14} = 
(s_{12})^{-2 - 2 \ep}
\left( -\frac{N_{\Gamma}}{b\,s\, x_1\, x_2\,} \right)
\,\times
\nonumber \\
&
 \Bigl[\,\log \left(\frac{1+\,s}{1-\,s}\right) \left(\log \left(\frac{-b\,s-x_{12\,}+2\,}{b-x_{12\,}+2\,}\right)+\log \left(\frac{b\,s-x_{12\,}+2\,}{b-x_{12\,}+2\,}\right)\right)
\nonumber \\
&
-\log \left(\frac{b\,s-x_{12\,}+2\,}{-b\,s-x_{12\,}+2\,}\right) \log \left(\frac{b+x_{12\,}-2\,}{b-x_{12\,}+2\,}\right)+\log(x_1)\, \times 
\nonumber \\
&
\left(\log \left(\frac{b\,s-x_{12\,}+2\,}{-b\,s-x_{12\,}+2\,}\right)+\log \left(\frac{b\,s+x_{12\,}+2\,}{-b\,s+x_{12\,}+2\,}\right)\right)
+\log \left(\frac{b\,s+x_{12\,}+2\,}{-b\,s+x_{12\,}+2\,}\right) \,\times
\nonumber \\
&
\left(-\log (x_2)-2\, \log \left(\frac{4
x_{12\,}}{x_{12\,}^2\,+4}\right)\right)
-2\, \log \left(\frac{-b\,s+x_{12\,}+2\,}{-b+x_{12\,}+2\,}\right)
\log \left(1-s\right)
\nonumber \\
&
+2\, \log \left(\frac{b\,s+x_{12\,}+2\,}{b+x_{12\,}+2\,}\right) \log \left(1-s\right)
-2\, \log \left(\frac{-b\,s+x_{12\,}+2\,}{b+x_{12\,}+2\,}\right) \log
\left(s+1\right)
\nonumber \\
&
+2\, \log \left(\frac{b\,s+x_{12\,}+2\,}{-b+x_{12\,}+2\,}\right)\,
\log \left(s+1\right)-2\,
\text{Li}_2\,\left(\frac{b-b\,s}{b-x_{12\,}+2\,}\right)
\nonumber \\
&
+2\, \text{Li}_2\,\left(\frac{b+b\,s}{b-x_{12\,}+2\,}\right)
-2\, \text{Li}_2\,\left(\frac{1}{2\,}
(-b\,s-x_{12\,})\right)
+2\, \text{Li}_2\,\left(\frac{b\,s-x_{12\,}}{2\,}\right)
\nonumber \\
&
+2\, \text{Li}_2\,\left(\frac{-b-b\,s}{-b+x_{12\,}+2\,}\right)
-2\, \text{Li}_2\,\left(\frac{b\,s-b}{-b+x_{12\,}+2\,}\right)
+2\, \text{Li}_2\,\left(\frac{b-b\,s}{b+x_{12\,}+2\,}\right)
\nonumber \\
&
-2\, \text{Li}_2\,\left(\frac{b+b\,s}{b+x_{12\,}+2\,}\right)
+2\, \text{Li}_2\,\left(\frac{-b\,s+x_{12\,}
-2\,}{b+x_{12\,}-2\,}\right)
-2\, \text{Li}_2\,\left(\frac{x_{12\,}-b\,s}{x_{12\,}-b}\right)
\nonumber \\
&
-2\, \text{Li}_2\,\left(\frac{x_{12\,}-b\,s}{b+x_{12\,}}\right)
-2\, \text{Li}_2\,\left(\frac{b\,s+x_{12\,}-2\,}{b+x_{12\,}-2\,}\right)
+2\,\text{Li}_2\,\left(\frac{b\,s+x_{12\,}}{x_{12\,}-b}\right)
\nonumber \\
&
+2\,\text{Li}_2\,\left(\frac{b\,s+x_{12\,}}{b+x_{12\,}}\right)
+ \mathcal{O}(\,\ep)
\;\Bigr]\,.
\end{align}

%%%%%%%%%%%%%%%%%%%%%%%%%%%%%%%%%%%%%%%%%%%%%%%%%%%%%%%%%%%%%%%%%%%%%%%%%%%%
where
\begin{align}
  \nonumber
  x_{12} &= x_1 + x_2\,, \\\nonumber
  b &= \sqrt{x_{12}^2+4}\,, \\\nonumber
  s &= \sqrt{Q_3/x_1}/b \,,\\ \nonumber
  Q_3 &= x_1\,(x_{12}^2 + 4) - 4\,x_{12}\,.\\ \nonumber
\end{align}
The master $I_{14}$ was calculated using differential equations and is 
written in terms of logarithms and dilogorithms of complicated arguments
instead of GHPL's and HPL's.
The coefficients of the differential equations for $I_{14}$ involve
a new denominator, cubic in $x_1$ and quadratic in $x_2$,
\begin{equation}
\label{eq:Q3def}
Q_3 = x_1 \left(x_1 + x_2\right)^2 - 4\,x_2
\end{equation}
which does not occur in the differential equations for any of the
other masters presented in this paper. Because of this denominator,
$I_{14}$ has this more complicated form.

The differential equations for $I_{14}$ are solved
by first considering the corresponding homogeneous equations.
A solution of these equations, for $\ep=0$, is given by
\begin{equation}
I_{14}^{\text{hom}} = \frac{1}{x_2 \sqrt{x_1} \sqrt{Q_3}}
\end{equation}
Note that it is singular along the curve $Q_3 = 0$, part
of which is {\em inside} the physical region
$0 < x_1 < 1, 0 < x_2 < 1$. On the other hand, the
full solution for $I_{14}$, which we write as
\begin{equation}
\label{eq:C14def}
I_{14} = C_{14} \, I_{14}^{\text{hom}}
\end{equation}
cannot be singular inside the physical region. Therefore,
the coefficient $C_{14}$ must vanish along the curve $Q_3 = 0$.
%By substituting eq.(\ref{eq:C14def}) into the differential equation for
%$I_{14}$, one finds that the derivatives of $C_{14}$ are given by
%\begin{eqnarray}
%\left( \frac{\partial  C_{14}}{\partial x_1} \right)_{x_2} & = &
%\frac{S_{\Gamma} }{x_2 \sqrt{x_1} \sqrt{Q_3}}
%\Biggl\{
%   \frac{(\ldots)}{(\ldots)}\, \log(x_1)
% \nonumber \\ &&
%{} + \frac{(\ldots)}{(\ldots)}\, \log(x_2)
%\nonumber \\ &&
%{} + ( 3\, x_1 + x_2 ) \bigl(  - \log(1+x_1) + \log(x_1+x_2) \bigr)
%\Biggr\}
%\\
%\left( \frac{\partial  C_{14}}{\partial x_2} \right)_{x_1} & = &
%\frac{S_{\Gamma} \sqrt{x_1}}{x_2 \sqrt{Q_3}}
%\Biggl\{
%  \frac{ x_1^2 + 2\, x_2 + 3\, x_1 x_2 + 2\, x_2^2 }{x_1+x_2}\, \log(x_1)
%\nonumber \\ &&
%{} + \frac{x_2 (x_1+x_2-2)}{x_1+x_2} \log(x_2)
%\nonumber \\
%&&
%  {} + (x_1-x_2) \bigl( \,\log(1+x_1) - \log(x_1+x_2) \bigr)
%\Biggr\}
%\, .
%\end{eqnarray}
In order to solve the differential equations for $I_{14}$, it is convenient to
consider them keeping $x_{12} = x_1 + x_2$ fixed, so that
$Q_3 = x_1 (x_{12}^2 + 4) - 4\,x_{12}$ becomes linear in $x_1$.
Using the boundary condition that $C_{14}$ must vanish at
the point $x_1 = x_1^0 = 4 x_{12}/(x_{12}^2+4)$, where $Q_3 = 0$,
the solution of the differential equation of $C_{14}$ in $x_1$ 
can then be written as a one-dimensional integral:
\begin{eqnarray}
%C_{14} &=& 
%\int_{x_1^0}^{x_1} \text{d} x'_{1}
% \left( \frac{ \parintial C_{14} }{ \partial x_1} \right)_{x_{12}}
%\nonumber \\
% MR 10.1.2011: replaced S_{\Gamma} by N_{\Gamma} because S_Gamma 
% does not show up anywhere
C_{14}       &=& \int_{x_1^0}^{x_1} \text{d} x'_{1}
 \frac{N_{\Gamma}}{\sqrt{x'_{1}}\sqrt{x'_{1}(x_{12}^2 + 4) - 4\,x_{12}}}
\nonumber \\
& \times & \Biggl\{
\left( \frac{-2 (2-x_{12})}{1-x'_1}
     + \frac{x_{12}^2}{x'_1-x_{12}}
     + \frac{2+x_{12}}{1+x'_1} \right) \log(x'_1)
\nonumber \\
&&
  {} + \frac{x_{12}^2}{x'_1-x_{12}}
       \bigl( \,\log(1+x'_1) - \log(x_{12}) \bigr)
     + \frac{2+x_{12}}{1+x'_1} \log(x_{12}-x'_1)
\Biggr\}
\end{eqnarray}
The integration can be performed analytically in terms of logarithms
and dilogarithms. The final result for $I_{14}$ is shown in eq.~(\ref{M14hard}).
\\\\
We have presented the result for $I_{14}$ in a form where all
(di)logarithms are real in the region where $Q_3 > 0$ and $s$ is real. By
construction, $I_{14}$ has no singularity at $Q_3 = 0$. For $Q_3 < 0$,
$s$ is imaginary and the arguments of the (di)logarithms that depend on
$s$ become complex. Nevertheless, $I_{14}$ remains real in this region
because the factor $C_{14}$ is now imaginary.
\\\\
Finally, it is worth noting that $I_{14}$ is only needed up to finite order in the hard 
region. It is required up to order $\epsilon$ in the collinear $x_1$ region 
where it has been calculated separately using differential equations and direct 
integration.

\subsubsection*{b) The collinear regions}
\label{sec:coll}
The real-real master integrals were calculated in the collinear regions
by deriving a hypergeometric integral representation starting from their
definition as phase space integrals. This allows us to give the results
in closed form. We also have derived these results as an expansion in $\ep$
using the differential equations method, which provides an important check
on our results. We list below the results of the masters contributing to
the antennae $\mathcal{B}_4^0,\;\mathcal{\tilde{E}}_4^0,\;
\mathcal{H}_4^0$ in these regions.
\begin{itemize}
\item {\bf \underline {collinear $x_1$}}
\end{itemize}
In this region,
only $I_{1}$, $I_{2}$, $I_{7}$ and $I_{8}$ require to be expanded 
up to ${\cal O}(\ep^3)$  since they contribute to $\tilde{E}_{12}$ which 
starts its power expansion at ${\cal O}(\ep^{-3})$.

\begin{eqnarray}
% MR 10.1.2011: k_a -> k_j, k_b -> k_k
%%%%I1
&&\I{1} \times k_k \cdot p_1\;\;=
I_1 =  
(s_{12})^{1-2 \ep}
\, N_{\Gamma}\, 2^{-2-\ep} \,(1 - x_1)^{1 - 2 \,\ep}\,
(1 - x_2)^{2 - 2 \,\ep} \, x_2^{-\ep} \, \times
\\ \nonumber
&& 
\hspace{5.5cm}(1+x_2)^{-1 + \ep}
\; \frac{\Gamma(1-\ep)^2}{ \Gamma(2-2\,\ep)}
\\
%%%%I2
&&\I{1} \;\;=
I_2 = 
(s_{12})^{-2 \ep}
N_{\Gamma}\, 2^{-\ep} \,(1 - x_1)^{1 - 2 \,\ep}\,
(1 - x_2)^{1 - 2 \,\ep} \, x_2^{-\ep} \, (1+x_2)^{-1 + \ep} \,\times
\\\nonumber&&
\hspace{3.7cm}\,\frac{\Gamma(1-\ep)^2}{\Gamma(2-2\,\ep)}
\\
%%%%I3
&&\I{6} \times (-(k_j \cdot p_1)(k_j \cdot p_3)) \;\;=
I_3 = 
(s_{12})^{1-2 \ep}
N_{\Gamma}\, 2^{-3-\ep} \,(1 - x_1)^{1 - 2 \,\ep}  \, \times
\\ \nonumber
&&
\hspace{4.5cm}
(1 - x_2)^{1 - 2 \,\ep}\, x_2^{-\ep}(1+x_2)^{-1 + \ep}\, \times
\\ \nonumber
&&
\hspace{4.5cm}
\left(1 -3\,x_2 + 2\, x_2^2 \; 
 _2F_1\left( 1, 1-\ep; 2 - 2\,\ep; 1-x_2\right)\right)
\;\frac{\Gamma(1-\ep)^2}{\Gamma(2-2\,\ep)}\;
\\
%%%%I4
&&\I{6} \times k_j \cdot p_1\;\;=
I_4 = -
(s_{12})^{-2 \ep}
 \, N_{\Gamma}\, 2^{-1-\ep} \,(1 - x_1)^{1 - 2 \,\ep}
(1 - x_2)^{1 - 2 \,\ep} x_2^{-\ep} \,\times 
\\ \nonumber
&&
\hspace{4.5cm}
(1+x_2)^{-1 + \ep}\,\left(-1 + x_2 \; 
 _2F_1\left( 1, 1-\ep; 2 - 2\,\ep; 1-x_2\right)\right)
\;\frac{\Gamma(1-\ep)^2}{\Gamma(2-2\,\ep)}
\\
%%%%I5
&&\I{6} \times (-k_j \cdot p_3)\;\;=
I_5 = -
(s_{12})^{-2 \ep}
 \, N_{\Gamma}\, 2^{-1-\ep} \times
 \\ \nonumber
 &&
 \hspace{6cm}
 (1 - x_1)^{1 - 2 \,\ep}\,
(1 - x_2)^{1 - 2 \,\ep} \, x_2^{-\ep} \,
(1+x_2)^{-1 + \ep} \times
\\ \nonumber
&&
\hspace{6cm}
\left(-1 + x_2 \; 
 _2F_1\left( 1, 1-\ep; 2 - 2\,\ep; 1-x_2\right)\right)
\;\frac{\Gamma(1-\ep)^2}{\Gamma(2-2\,\ep)}
\\ 
%%%%I6
&&\I{6} \;\;=
I_6 =  
(s_{12})^{-1-2 \ep}
N_{\Gamma}\, 2^{-\ep} \,(1 - x_1)^{1 - 2 \,\ep}\,
(1 - x_2)^{1 - 2 \,\ep} \, x_2^{-\ep} \, \times
\\ \nonumber
&&  
\hspace{3.7cm}(1+x_2)^{-1 + \ep} \,\frac{\Gamma(1-\ep)^2}{\Gamma(2-2\,\ep)} \;
 _2F_1\left( 1, 1-\ep; 2 - 2\,\ep; 1-x_2\right)
\\
%%%%I7
&&\I{7} \;\;=
I_7 = -
(s_{12})^{-1-2 \ep}
 N_{\Gamma}\, 2^{-1-\ep} \,(1 - x_1)^{- 2 \,\ep}\,
(1 - x_2)^{1 - 2 \,\ep} \, x_2^{-1-\ep} \, \times
\\ \nonumber
&& 
\hspace{4cm}
(1+x_2)^{\ep} \, _2F_1\left( 1, 1-\ep; 2 - 2\,\ep; \frac{1-x_2}{2}\right)\;
\,\frac{\Gamma(1-\ep)^2}{\Gamma(2-2\,\ep)}
\\ 
%%%%I8
&&\I{8} \;\;=
I_8 = 
(s_{12})^{-1-2 \ep}
 \, N_{\Gamma}\, 2^{-1-\ep} \,(1 - x_1)^{- 2 \,\ep}\,
(1 - x_2)^{1 - 2 \,\ep} \,x_2^{-1-\ep} \, \times
\\ \nonumber
&& 
\hspace{4cm}
 (1+x_2)^{\ep} \, _2F_1\left( 1, 1-\ep; 2 - 2\,\ep; \frac{-1+x_2}{2\,x_2}\right)
\,\frac{1}{\ep} \frac{\Gamma(1-\ep)^2}{\Gamma(1-2\,\ep)}
\\
%%%%I14
&&\I{14} \;\;=
I_{14} = -
(s_{12})^{-2-2 \ep}
 N_{\Gamma}\, 2^{-1-\ep} \,(1 - x_1)^{- 2 \,\ep}\,
(1 - x_2)^{1 - 2 \,\ep} \,x_2^{-1-\ep} \, \times
\\ 
&& 
\hspace{2.5cm}
(1+x_2)^{\ep} \, _2F_1\left( 1, 1-\ep; 2 - 2\,\ep; \frac{1-x_2}{2}\right)\;
 _2F_1\left( 1, 1-\ep; 2 - 2\,\ep; 1-x_2\right)
\,\frac{\Gamma(1-\ep)^2}{\Gamma(2-2\,\ep)}
\nonumber
\end{eqnarray}

\begin{itemize}
\item {\bf \underline {collinear $x_2$}}
\end{itemize}
In this region, only the following four masters (all of them are needed up to ${\cal O}(\ep^2)$) 
contribute:
\begin{eqnarray}
% MR 10.1.2011: k_a -> k_j, k_b -> k_k
%%%%I1
&&\I{1}  \times k_k \cdot p_1 \;\;= 
I_1 = 
(s_{12})^{1-2 \ep}
N_{\Gamma}\, 2^{-3-\ep} \,(1 - x_1)^{1 - 2 \,\ep}\, \times
\\\nonumber
&&
\hspace{5.3cm}
(1 - x_2)^{2 - 2 \,\ep} \,  x_1^{-\ep} \, (1+x_1)^{-2 + \ep} \, (1 +3\,x_1)\; \frac{\Gamma(1-\ep)^2}{ \Gamma(2-2\,\ep)}
\\
%%%%I2
&&\I{1} \;\;= 
I_2 = 
(s_{12})^{-2 \ep}
N_{\Gamma}\, 2^{-\ep} \,(1 - x_1)^{1 - 2 \,\ep}\,
(1 - x_2)^{1 - 2 \,\ep} \, x_1^{-\ep} \, \times
\\\nonumber
&&
\hspace{4cm}
\, (1+x_1)^{-1 + \ep} \,\frac{\Gamma(1-\ep)^2}{\Gamma(2-2\,\ep)}
\\
%%%%I7
&&\I{7} \;\;= 
I_7 = -
(s_{12})^{-1-2 \ep}
N_{\Gamma}\, 2^{-\ep} \,(1 - x_1)^{ - 2 \,\ep}\,
(1 - x_2)^{1 - 2 \,\ep} \, x_1^{-\ep} \, \times
\\\nonumber
&&
\hspace{4.1cm}
\, (1+x_1)^{-1 + \ep} \, \frac{\Gamma(1-\ep)^2}{\Gamma(2-2\,\ep)}
\\\nonumber
%%%%I10
&&\I{10} \;\;= 
I_{10} =  -
(s_{12})^{-1-2 \ep}
N_{\Gamma}\, 2^{-1-\ep} \, (1 - x_1)^{1 - 2 \,\ep}\,
(1 - x_2)^{- 2 \,\ep} \, x_1^{-1-\ep} \, \times
\\
&& \hspace{4.3cm} 
\, (1+x_1)^{ \ep} \, \frac{\Gamma(1-\ep)^2}{\Gamma(2-2\,\ep)}\,
   _2F_1\left(1, 1 - \ep; 2 - 2 \,\ep; \frac{1 - x1}{2}\right)
\end{eqnarray}

\subsubsection*{c) The soft region}
\label{sec:soft}
As discussed previously, the master integrals in the soft region are
expected to be needed at most two orders in $\ep$ higher
than in the hard region. It turns out here  that the masters are only
needed up to ${\cal O}(\ep^3)$ at most in the soft region. The reason behind 
this is the absence of double soft gluon configurations in the antennae 
$B$, $E$ and $H$. 
Furthermore, as can be seen from Table 1, only $B_{12}$ will have a 
contribution from the soft region at all. This is due to the fact 
that only $B_{12}$ allows for a 
singular double soft configuration, a soft $q-\bar{q}$ pair.
The master integrals which are required in this case up to 
${\cal O}(\ep^3)$  are either $I_{2}$ or ${I}_{1}'$ depending 
on the basis choice for the masters. The others are only needed up 
to ${\cal O}(\ep^2)$.  

In the soft region, the masters are calculated by a direct
evaluation of the phase space integrals.
In this case, the  results are
products of gamma functions and therefore can be presented
in closed form.

\begin{eqnarray}
% MR 10.1.2011: k_a -> k_j, k_k -> k_k
%
%%%%I1
&&\I{1} \times k_k \cdot p_1 \;\;= 
\,I_1 = 
(s_{12})^{1-2\ep}
 N_{\Gamma}\,(1 - x_1)^{1 - 2 \,\ep}\,(1 - x_2)^{2 - 2 \,\ep} \,\times
\\\nonumber
&&
\hspace{5.5cm}
\frac{\Gamma(1-\ep)^2}{8 \,\Gamma(2 - 2 \,\ep)}
\\
%%%%I2
&&\I{1}\;\;= 
\,I_2 =  
% MR 24.9.2010 inserted s12
(s_{12})^{-2\ep}
N_{\Gamma}\,(1 - x_1)^{1 - 2 \,\ep}\,(1 - x_2)^{1 - 2 \,\ep} 
\,\,\frac{\Gamma(1-\ep)^2}{2 \,\Gamma(2 - 2 \,\ep)}
\\ 
%%%%I7
&&\I{7} \;\;= 
\,I_7 = -
% MR 24.9.2010 inserted s12 and inserted a line break
(s_{12})^{-1-2\ep}
N_{\Gamma}\,(1 - x_1)^{- 2 \ep}\,(1 - x_2)^{1 - 2 \,\ep} \, \times
\\ \nonumber
&&
\hspace{5.5cm} 
\frac{\Gamma(1-\ep)^2}{2 \,\Gamma(2 - 2 \,\ep)}
\\ 
%%%%I10
&&\I{10} \;\;= 
\,I_{10} = -
% MR 24.9.2010 inserted s12 and inserted a line break
(s_{12})^{-1-2\ep}
N_{\Gamma}\,(1 - x_1)^{1- 2 \,\ep}\,(1 - x_2)^{- 2 \,\ep} \, \times
\\ \nonumber
&&
\hspace{5.5cm}
\frac{\Gamma(1-\ep)^2}{2 \,(1-2\,\ep)\,\Gamma(1 - 2 \,\ep)} 
\end{eqnarray}

\subsubsection{Integrated antennae}
In this section, we present the results for the integrated 
crossings of three final-final four parton antenna functions 
$B_4^0(q,q',\bar{q}',\bar{q})$,
$\tilde{E}_4^0(q,q',\bar{q}',g)$
and $H_4^0(q,\bar{q},q',\bar{q}')$ defined in \cite{GehrmannDeRidder:2005cm}.
Because the results are too long to present fully here, we show
explicitly only the pole terms (all pole terms higher than
$1/\ep^2$, and only the highest pole term if it
is $1/\ep$). The dimensionful prefactor $(s_{12})^{-2\ep}$ 
of every integrated antenna is omitted below. The complete results are included
in a {\tt Mathematica} file appended to the source file of the manuscript.

The integrated form of the initial-initial antenna functions
$H_{12} $ and $H_{13} $ are obtained by first crossing a pair 
of identical or non-identical quarks
in the final-final antenna $H_4^0(1q,2\bar{q},3q',4\bar{q}')$
to the initial state, and then calculating the phase space integrals
as described in the previous sections. They take the following form:
\begin{eqnarray}
\mathcal{H}_{12} &=& - \frac{1}{\ep}\;\Bigl\{\frac{(x_1 x_2\,+1)\, \left(\left(x_2^4+\left(x_1^2-4\right)\,
  x_2^2+1\right)\, x_1^2+x_2^2\right)\,}{3 \, (x_1+x_2\,)^4}\Bigr\} + \order{1} \,,
\\
\mathcal{H}_{13} &=& \,\frac{1}{\ep^2}\; \Bigl\{\frac{\left(x_1^2-2 x_1+2\right)
\left(x_2^2-2 x_2+2\right)}{4 \,x_1 x_2}\Bigr\} + \order{ \frac{1}{\ep} } \,.
\end{eqnarray}

For the $\tilde{E}_4^0(1q,2q',3\bar{q}',4g)$ antenna, there are
four independent expressions, obtained by crossing 
$qq'$, $qg$, $q'\bar{q}'$ or $q'g$ to the initial state,
we list the pole terms of the integrated ones:
\begin{eqnarray}
 \mathcal{\tilde{E}}_{12} &=& \frac{1}{\ep^3}\;\Bigl\{\frac{\left(x_2^2-2 x_2+2\right)\,
 \delta (1-x_1)\,}{4 \,x_2}\Bigr\}
\\\nonumber
     &+&\frac{1}{\ep^2}\;\Big\{
     -\frac{1}{2 (x_1+1)\, x_2
     (x_1+x_2)^2} \left( -2 x_1^2 x_2^4-2 x_1 x_2^4- x_2^4 +2 x_1 x_2^3 \right. 
\\\nonumber&&
	\left. \qquad  \qquad +2 x_2^3+x_1^2 x_2^2-2 x_1 x_2^2-2
    	 x_2^2-2 x_1^2 x_2+2 x_1^2 -4 x_1 x_2 +4 x_1 \right)
\\\nonumber &&
    -\frac{\left(x_2^2-2 x_2+2\right)\,
     \mathcal{D}_0(x_1)\,}{2 x_2}
     +\delta (1-x_1)\, \left(\frac{\left(x_2^2-2 x_2+2\right)\, H(-1,x_2)\,}{2 x_2}
\right. \\\nonumber && \left.
     -\frac{\left(3 x_2^2-6 x_2+8\right)\, H(0,x_2)\,}{8 x_2}
     +\frac{\left(x_2^2-2 x_2+2\right)\, H(1,x_2)\,}{2 x_2}
\right. \\\nonumber && \left.
     -\frac{8 \log (2)\, x_2^2-7 x_2^2-16 \log (2)\, x_2+24 x_2+16 \log (2)\,-20}
     {16 x_2}\right)\,\Bigr\} + \order{ \frac{1}{\ep} } \,,
\\
\mathcal{\tilde{E}}_{14} &=& 
\frac{1}{\ep^2}\;\Bigl\{\delta (1-x_1) \left(\frac{(x_2+1) H(0,x_2)}{2}-\frac{(x_2-1) 
\left(4 x_2^2+7 x_2+4\right)}{12 x_2}\right)\Bigr\} + \order{ \frac{1}{\ep} } \,
\\
\mathcal{\tilde{E}}_{23} &=& \frac{1}{\ep^2} \; \Bigl\{\frac{x_1 x_2 (x_1 x_2+1)^2
  \left(x_1^2+x_2^2-2\right)} {(x_1+x_2)^4} \Bigr\} + \order{ \frac{1}{\ep} }
 \,,
\\
\mathcal{\tilde{E}}_{24} &=&
\frac{1}{\ep}\;\Bigl\{- \frac{1}{6 x_1^3 (x_1+x_2)^3}\Bigl[(x_2-1)
  \left(12 x_1^9+36 x_2 x_1^8-6 x_1^8+41
  x_2^2 x_1^7 -19 x_2 x_1^7
\right. \\\nonumber && \left.
  -x_1^7
  +19 x_2^3 x_1^6-29 x_2^2 x_1^6-5 x_2 x_1^6+4
  x_2^4 x_1^5-20 x_2^3 x_1^5-26 x_2^2 x_1^5+15 x_2 x_1^5
\right. \\\nonumber && \left.
  -6 x_2^4 x_1^4
  -30 x_2^3 x_1^4+73 x_2^2 x_1^4+10 x_2 x_1^4-2 x_1^4-12 x_2^4 x_1^3+86 x_2^3
  x_1^3+50 x_2^2 x_1^3
\right. \\\nonumber && \left.
 -34 x_2 x_1^3
  +32 x_2^4 x_1^2+62 x_2^3 x_1^2-118 x_2^2
  x_1^2+24 x_2^4 x_1-132 x_2^3 x_1-48 x_2^4\right)\Bigr]
\\\nonumber&&
+\frac{\left(2 x_1^8+x_2 x_1^7-x_1^6+2 x_2^2 x_1^4-x_2 x_1^3-8 x_2^2 x_1^2+2
  x_2 x_1+8 x_2^2\right)}{x_1^4 } \,\times
\\\nonumber &&
  \left(G(-x_2,x_1) - H(-1,x_1) \right)
  + x_1^2 \, \left( 2 x_1^2 + x_1 x_2 - 1 \right) H(0,x_2) \Bigr\} + \order{1} \,.
\end{eqnarray}
Finally, for the $B_4^0(1q,3q',4\bar{q}',2\bar{q})$
antenna, we have three independent crossings, obtained by crossing
either the primary quarks $q\bar{q}$, the secondary ones $q'\bar{q}'$,
or a combination of primary and secondary quarks to the initial state.
Below we show the pole terms of $B_{12}$ and $ B_{13}$,
the antenna $B_{34}$ is completely finite.
\begin{eqnarray}
\mathcal{B}_{12} &=&
- \frac{1}{\ep^3}\;\Bigl\{\frac{\delta (1-x_1) \,\delta (1-x_2)}{12}\Bigr\}
\\\nonumber
&&+\frac{1}{\ep^2}\;\Bigl\{\delta(1-x_1) \left(-\frac{1+x_2}{12}+\frac{1}{6}
  \mathcal{D}_0(x_2)-\frac{5}{36} \delta (1-x_2)\right)
\\\nonumber&&
  +\left(\frac{1}{6}
  \mathcal{D}_0(x_1)-\frac{1+x_1}{12}\right) \delta 
  (1-x_2)\Bigr\} + \order{ \frac{1}{\ep} },
\\
\mathcal{B}_{13} &=& \frac{1}{\ep^2} \delta(1-x_1) \;\Bigl\{
\frac{ (1-x_2) (4 x_2^2 + 7 x_2 + 4 ) }{ 24 x_2 } + \frac{1+x_2}{4} H(0, x_2) \Bigr\}
\\\nonumber&&
+ \order{\frac{1}{\ep}} \,.
\end{eqnarray}

\section{Conclusions}
Within the antenna subtraction formalism, allowing the calculation 
of higher order QCD corrections to jet observables, subtraction terms 
are constructed from antenna functions. Those functions describe 
all unresolved radiation between a pair of hard radiator partons. 
At NNLO, this formalism has been fully developed and applied 
so far only for colourless initial states.
In this paper, we have focussed on the extension of this formalism to evaluate 
NNLO corrections to jet observables at hadron colliders and concentrated 
on the construction of subtraction terms for the double real radiation 
contributions.  
More precisely, we have considered the subtraction terms needed to account 
for the radiation of two colour-connected unresolved partons off two initial
state partons. For these subtraction terms, 
four-parton tree-level initial-initial antenna functions are required
in unintegrated and integrated form. 
The integration over the phase space associated with two unresolved partons 
has to be performed analytically.
 
In this paper, we have given a catalogue of all non-identical 
four-parton initial-initial antenna functions. Furthermore, after 
applying standard reduction techniques, we found that 32 master 
integrals are necessary to obtain their integrated form. 
As a step towards the integration of the full set of integrated initial-initial antenna
functions, in this paper we have focussed on the initial-initial antennae
obtained from crossing two partons in the final-final antenna functions 
characterised by the presence of two quark flavours.
After reduction, 12 masters were required to obtain those.
We presented the decomposition of the calculation according to four phase space
regions: hard, collinear and soft and we gave the master integrals 
in these regions. Finally, we presented the results for those integrated 
initial-initial four parton antenna functions themselves.
Since the results are lengthy, we have shown only the leading pole terms in
the manuscript and have attached the complete results as
a { \bf \tt Mathematica }file.

\section{Acknowledgements}
This work is supported in part by the U.S. Department of
Energy, Division of High Energy Physics, under contract
DE-AC02-06CH11357 and by the Swiss National Science Foundation (SNF)
under contract PP0022-118864.

\bibliographystyle{JHEP-2}

\providecommand{\href}[2]{#2}\begingroup\raggedright\endgroup

\end{document}